\documentclass[11pt]{article}

\usepackage[table]{xcolor}
\usepackage{amsmath}
\usepackage{amssymb} 
\usepackage{graphicx,psfrag,epsf}
\usepackage{enumerate}
\usepackage[round,sort&compress,authoryear]{natbib}
\usepackage[hyphens]{url} 
\usepackage{ulem} 
\usepackage{standalone}
\usepackage[justification=justified,singlelinecheck=false]{caption}
\usepackage{caption}
\DeclareCaptionLabelFormat{hyphen}{#1-#2}
\usepackage{booktabs}
\usepackage{multirow}
\usepackage{float} 
\newlength{\mcol}
\newlength{\lcol}
\newcommand{\blind}{0}

\usepackage{makecell}
\usepackage{amsthm}
\usepackage{bm}
\usepackage{bbm}
\usepackage{mathtools}
\usepackage{collcell}
\usepackage{dsfont}
\usepackage{rotating}
\usepackage{wrapfig}
\usepackage{tabularx}
\usepackage[utf8]{inputenc}
\usepackage{tikz}
\usepackage{lettrine}
\usepackage{dblfloatfix}
\usepackage[english]{babel} 
\usepackage{amsfonts}
\usepackage{animate}
\usepackage{subcaption}
\usepackage{floatflt}
\graphicspath{ {fig/} }
\usepackage[hyperfootnotes=false]{hyperref}

\usepackage[justification=centering]{caption} 
\usepackage{arydshln}
\usepackage{lipsum}
\usetikzlibrary{arrows,positioning,decorations.markings}
\usepackage{standalone}
\usepackage{tablefootnote}
\usepackage{booktabs}
\usepackage{float}

\usepackage{etoolbox}
\makeatletter
\def\tikz@valign{c}
\tikzset{
  enforce alignment/.code={
    \csname if#1\endcsname
      \def\tikz@text@width
        {\pgfkeysvalueof{/pgf/minimum width}-2*(\pgfkeysvalueof{/pgf/inner xsep})}%
    \else
      \let\tikz@text@width\pgfutil@empty
    \fi},
  enforce alignment/.default=true,
  valign/.is choice,
  valign/top/.code=\def\tikz@valign{t},
  valign/center/.code=\def\tikz@valign{c},
  valign/bottom/.code=\def\tikz@valign{b},
  valign height/.initial=%
    \pgfkeysvalueof{/pgf/minimum height}-2*(\pgfkeysvalueof{/pgf/inner ysep})}
\patchcmd\tikz@fig@continue{\tikz@node@transformations}{%
  \pgfmathsetlength\pgf@x{\pgfkeysvalueof{/tikz/valign height}}%
  \pgf@y\ht\pgfnodeparttextbox
  \advance\pgf@y\dp\pgfnodeparttextbox
  \ifdim\pgf@y<\pgf@x
  \if\tikz@valign b%
    \advance\pgf@x-\dp\pgfnodeparttextbox
    \setbox\pgfnodeparttextbox\vbox to \pgf@x{\vfill\hbox{\unhbox\pgfnodeparttextbox}}%
  \else\if\tikz@valign t%
    \setbox\pgfnodeparttextbox\vbox to \pgf@x{\hbox{\unhbox\pgfnodeparttextbox}\vfill}%
  \fi\fi\fi
  \tikz@node@transformations}{}{}
\makeatother

\definecolor{gray}{rgb}{0.5,0.5,0.5}
\definecolor{red}{rgb}{0.8,0,0}
\definecolor{dred}{rgb}{0.5,0,0}
\definecolor{blue}{rgb}{0,0.1,1}
\definecolor{dblue}{rgb}{0,0.1,0.6}
\definecolor{cyan}{rgb}{0,0.5,.5}
\definecolor{dcyan}{rgb}{0,0.3,.3}
\definecolor{mpurple}{rgb}{.7,0,.9}

\definecolor{b}{rgb}{0,0,.8}	
\definecolor{g}{rgb}{0,.6,0}	
\definecolor{n}{rgb}{0,0,0}	
\definecolor{h}{rgb}{0.4,0.2,0.2}	
\definecolor{v}{rgb}{0.2,0.6,0}

\newcommand{\CC}{{\mathcal{C}}}

\newcommand{\FF}{{\mathcal{F}}}

\newcommand{\PP}{{\mathcal{P}}}

\newcommand{\ov}\overline

\newcommand{\rig}\right
\newcommand{\lef}\left
\newcommand{\nf}\normalfont

\usepackage{forest} 
\usepackage{multirow}

\usepackage{longtable}

\setcitestyle{authoryear,comma,aysep{,}}

\begin{document}

	\def\spacingset#1{\renewcommand{\baselinestretch}%
		{#1}\small\normalsize} \spacingset{1}

	
	\if0\blind
	{
		\title{\bf Networked Spatial Effects in European Electricity Price Forecasting}
		\author{Sultan Mahmud Chomon,\footnote{Corresponding author.} \footnote{Chair of Data Science in Energy and Environment, House of Energy Markets and Finance, University of Duisburg-Essen, Essen, Germany. Email addresses: sultan.chomon@uni-due.de (Sultan Mahmud Chomon), florian.ziel@uni-due.de (Florian Ziel).} , Florian Ziel\footnotemark[2] \\
		University of Duisburg-Essen, Germany}
		\maketitle
	} \fi

	\if1\blind
	{
		
		\bigskip
		
		\begin{center}
			{\LARGE\bf Spatial Effects}
		\end{center}
	
	} \fi


	\begin{abstract} \noindent
    As European bidding zones are highly interconnected by physical transmission lines, spatial influences propagate across neighboring nodes through a network. It is reflected in the day-ahead electricity prices across European bidding zones, as the auction algorithm also uses information beyond each bidding zone's geographic boundary. To capture how this interconnection affects the neighboring bidding zone’s electricity prices, we have used a metric graph to map the spatial coverage of information using a well-defined neighborhood measure. We propose the Networked Spatio-Temporal Model (NSTM), which maps irregular spatial nodes into an ordered network, enabling the systematic incorporation of neighborhood information. We implement the NSTM across 39 bidding zones covering the majority of European electricity markets in a high-resolution, streaming-forecasting setup. The model uses autoregressive, cross-hour, and seasonal effects, along with fuel and emission prices and day-ahead forecasts of fundamentals, as interconnected information to predict the day-ahead prices for each bidding zone. A Europe-wide study presented in this paper shows that the NSTM consistently outperforms traditional island-based pure local models. This paper provides a framework that demonstrates the critical role the networked structure plays in propagating information across interconnected markets and its vast implications on day-ahead electricity price forecasting.

	\end{abstract}
	
	\noindent%
	\vfill
	
	\spacingset{1.45} 

\newpage
\section{Introduction} The Day-Ahead electricity price auction is the cornerstone of the European electricity markets. Cross-border electricity flows are highly significant across European electricity markets. Naturally, it deeply affects the Day-Ahead (DA) electricity prices of trading partners. Due to the Europe-wide market coupling mechanism, the spatial effects across bidding zones are already playing an important role, and the relevance is increasing day by day. The current electricity price forecasting literature primarily focuses on bidding zone-specific information and thus often fails to account for spillover effects from neighboring bidding zones. In contrast, we demonstrate the importance of networked spatial structures for European electricity price forecasting that goes beyond the geographic boundaries of the bidding zone under consideration. The meta-study by \cite{cb1_deblauwe2025} reviews cross-border effects and discusses their relevance for electricity price formation. The study finds that the central Western European markets are highly integrated and share common drivers, and that the largest cross-border effects occur in small zones near major renewable generation centers. \vspace{.2cm} 

The electricity price process is displays complex seasonality, and balances supply and demand within physical constraints. The price mechanism is governed by both the unique traits of bidding zones and network spillover in the larger system. Many factors may contribute to fostering a degree of homogeneity across the European bidding zones. The similarities in European economic activities, meteorological conditions, geography, and seasonal patterns result in quite synchronized demand changes across Europe. Strong evidence for the impact of market integration or coupling in European day-ahead electricity price forecasting is documented in (\cite{MILago2018}, \cite{Varga2025}, \cite{ZIEL2015430}, \cite{mi_karahan2024}). \noindent Additionally, mechanisms such as the Single Day-Ahead Market Coupling (SDAC) and flow-based market coupling in the Core Capacity Calculation Region are becoming increasingly dominant in the European pricing mechanism. With cross-border electricity trade, power generators must offer fair prices to compete with neighboring price zones. Taken together, these factors suggest that cross-bidding zone information may be highly influential, if not essential, for accurately capturing the Day-Ahead electricity price process. 

\begin{figure}[H]
    \centering
    \captionsetup{justification=justified,singlelinecheck=false, font=footnotesize}
    \includegraphics[width=1.0\textwidth]{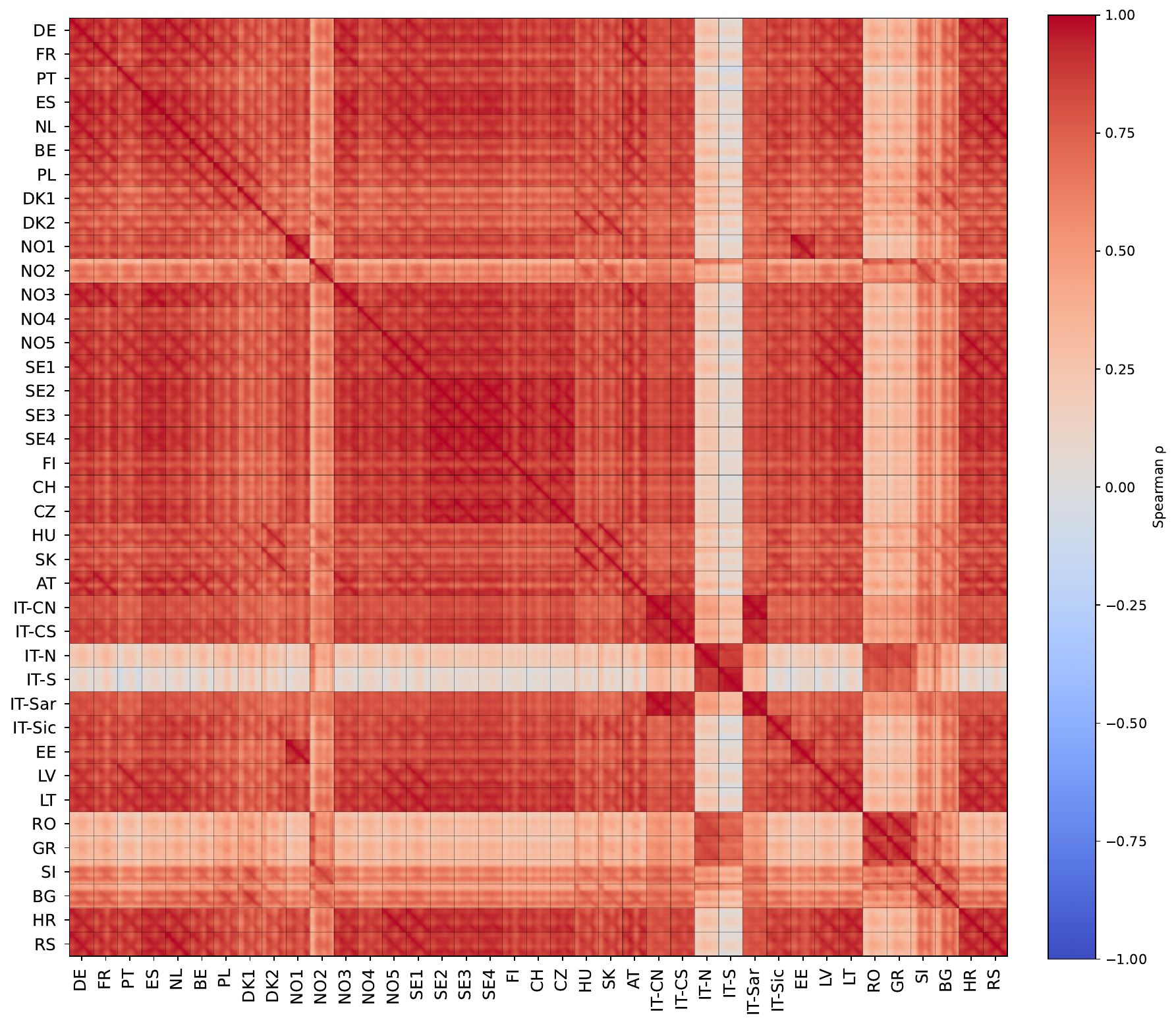}
    \caption{The spatio-temporal correlations of day-ahead electricity prices during the training period (Oct 2018–Sep 2023) are shown. These correlations are estimated using the Spearman rank correlation coefficient,
\[
\rho_{ij} =
\frac{\sum_{d=1}^{D} (R_{i,d}-\bar{R}_i)(R_{j,d}-\bar{R}_j)}
{\sqrt{\sum_{d=1}^{D} (R_{i,d}-\bar{R}_i)^2}\sqrt{\sum_{d=1}^{D} (R_{j,d}-\bar{R}_j)^2}},
\]
Here, $R_{i,d}$ and $R_{j,d}$ represent the ranks of electricity prices object $i$ and $j$ on day $d$. Within each block, a $24 \times 24$-hour matrix displays hourly dynamics. Off-diagonal blocks capture correlations between different bidding zones and hours.
}
\label{fig:winter-zoneblocks}
\end{figure}

\citet{Schnabel2025} argues that Europe-wide increases in renewable penetration and cross-border flows will heighten the importance of accounting for simultaneous interactions across multiple bidding zones. Using a panel data approach across 12 bidding zones, the paper shows that cross-border information benefits forecasting performance. \citet{cb_Stiewe2025} studied 30 European BZNs and found that neighboring BZNs' wind and solar energy outputs are highly influential for domestic renewable energy prices. Similarly, using space-time modeling, \citet{abate2017space} found significant spatial dependence in  spot price dynamics in the Nordic region. Renewable integration has become increasingly prominent, and geopolitical risk in energy markets has crystallized in European economies, making both of them vital. The spill-over has been exacerbated by the interconnected European economies and the interaction among the regressor classes. For instance, an interconnected market with high renewable generation interacts with natural gas and carbon markets.  For more, see \citet{CBDo2024}. In a different context, in New Zealand, \citet{Wen2022_wind_hydro_nz} observed that wind has a substantial downward spillover impact on prices and increased volatility. To address the growing complexity in spatial design, machine learning and deep learning methodologies are gaining popularity in the literature. For example, \citet{CBYang2024} proposes leveraging a deep learning methodology (a Graph Neural Network) to capture the complex spatio-temporal relationship and finds it effective for improved forecasting. \citet{cb3_aliyon2024} shows that a deep learning-based model (a daily-recalibrating multilayer perceptron) can be useful for uncovering the relationship between price and volatility.

\begin{table}[htbp]
\caption{Literature review of spatial aspects on EPF}
\centering
\fontsize{8.5}{8.5}\selectfont
\label{tab:epf_studies}

\begin{tabularx}{\textwidth}{
>{\raggedright\arraybackslash}p{2.2cm}
>{\raggedright\arraybackslash}p{1.5cm}
>{\raggedright\arraybackslash}p{2cm}
>{\raggedright\arraybackslash}p{2.8cm}
>{\raggedright\arraybackslash}X
}
\hline
Paper & Period & Market & Models & Scope of the study \\
\hline
\cite{cb_mascarenhas2026} & 2022--2024 & Belgian and Swedish BZNs & Statistical models & Cross-border asynchronous information significantly improves model performance during 2024 \\ \hline
\cite{cb_Stiewe2025} & 2015--2023 & 30 EU BZNs & Panel data econometric models & Cross-border spillover effect of wind and solar on EU electricity prices \\ \hline
\cite{CBDo2024} & 2012--2022 & 11 EU countries & SStatistical models & Cross-border spillover effects in European electricity markets \\ \hline
\cite{CBTrebbien2024} & 2019--2023 & Nordic and the Iberian Peninsula & Statistical (PCA) & Spatio-temporal aspects in electricity price formation \\ \hline
\cite{CBYang2024} & 2013--2018 & 10 EU BZNs (Nordics and Baltics) & Machine/Deep Learning (Graph Neural Network) & Spatio-temporal impacts on neighbouring DA electricity prices \\ \hline
\cite{Schnabel2025} & 2019--2024 & 13 EU BZNs & Panel data econometric model & Cross-border flows, renewable penetration impacts on neighbouring DA electricity prices \\ \hline
\cite{Wen2022_wind_hydro_nz} & 2011--2012 & New Zealand & Panel data econometric models & Wind and hydro impacts on electricity prices in a seasonal spatial econometric setup \\ \hline
\cite{cb3_aliyon2024} & 2015--2023 & 19 European BZNs & Multilayer perceptron & Use of deep learning for electricity price forecasting \\ \hline
\cite{CBbille2023} & 2015--2019 & Italian BZNs & Statistical/time series econometrics & Analysis of linear and non-linear models with an expert-type model with cross-border effects \\ \hline
\cite{CBAbrell2022} & 2015--2020 & Germany & Statistical/time series & Cross-border, merit order effects on electricity prices \\ \hline
\cite{CBMadadkhani2024} & 2016--2021 & Germany & Machine learning & Impact of fuel class along with cross-border power market features \\ \hline
\cite{CBMacedo2021} & 2016--2020 & Sweden & Statistical/time series econometrics & Impact of cross-border import-export and merit order on electricity prices \\ \hline
\cite{CB_Keles2020} & 2011--2017 & Switzerland & Fundamental (Nash-Cournot model) and statistical model & Cross-border effects on interconnected electricity market \\ \hline
\cite{CBAbadie2021} & 2016--2019 & France and Spain & Statistical and fundamental models & Cross-border effects on interconnected electricity market \\ \hline
\cite{CB_AnnanPhan2018} & 2012--2014 & Switzerland & Time series models & Cross-border effects on Swiss electricity prices \\ \hline
\cite{CBFrauendorfer2018} & 2012--2014 & 13 Nordic BZNs & Panel data econometric model & Spatial dependence on DA electricity prices \\ \hline
\cite{abate2017space} & 2012--2014 & 13 Nordic BZNs & Panel data econometric model & Spatial dependence on DA electricity prices \\ \hline
\end{tabularx}
\end{table}
\vspace{.2cm}

Distinct local patterns across interconnected European power grids complicate the estimation and integration of a global unified spatial process. The case of spatial non-stationarity stems from regional fragmentation, variations in power park characteristics, subregional similarities or differences in geography and meteorology, transmission bottlenecks, and uneven renewable energy adoption. Furthermore, merit order curves for each bidding zone respond asymmetrically to energy input prices, such as power plant fuel and emission prices, as well as to changes in the generation mix. This creates unique behavioral patterns by bidding zone. Since electricity demand is highly seasonal, seasonality influences the shape of the electricity price trajectory. While spatial relationships are crucial, the following are some complications that make simultaneous incorporation of spatio-temporal dimensions much harder: (1) Spatio-temporal relationships are complex, high-dimensional, and potentially nonlinear. (2) Design challenges in the modeling of the spatial components. (3) Computational overhead rises with large bidding zones and features used. Deep learning methods are effective but introduce significant computational overhead, and the underlying mechanisms remain opaque and difficult to interpret. \vspace{.2cm} 

In the existing literature, EPF modeling options are quite broad. \citet{weron2014_review} categorized the modeling approaches into three groups: Multi-Agent, fundamental, reduced-form, statistical, and computational intelligence models. In the meta-study by \cite{cb1_deblauwe2025}, they classify 3 classes of model researchers used to capture the cross-border effects: (1) Statistical models (Univariate and multivariate), (2) Fundamental market models, and (3) Explainable ML/AI methods. The fundamental model is highly useful in policy analysis, whereas the statistical and ML/AI models are quite reasonable choices for sharper short-term forecasts. The statistical models can leverage fundamental characteristics to achieve sharper price forecasting as such in \citet{UNIEJEWSKI2026125844}. Also, The combination of fundamental and statistical models are found to be useful for better forecasting performance. For more. see \citet{Paulmo206}. ML/AI models are quite successful at modeling complex nonlinear relationships but are often computationally very costly and black-box, with highly sensitive hyperparameters. On the other hand, statistical models offer greater explanatory power and transparency and often strike a good balance between accuracy and cost. We have chosen to pursue the latter. \vspace{.2cm}

\begin{figure}[H]
    \centering
    \captionsetup{justification=justified,singlelinecheck=false, font=footnotesize}
    \includegraphics[width=0.9\textwidth]{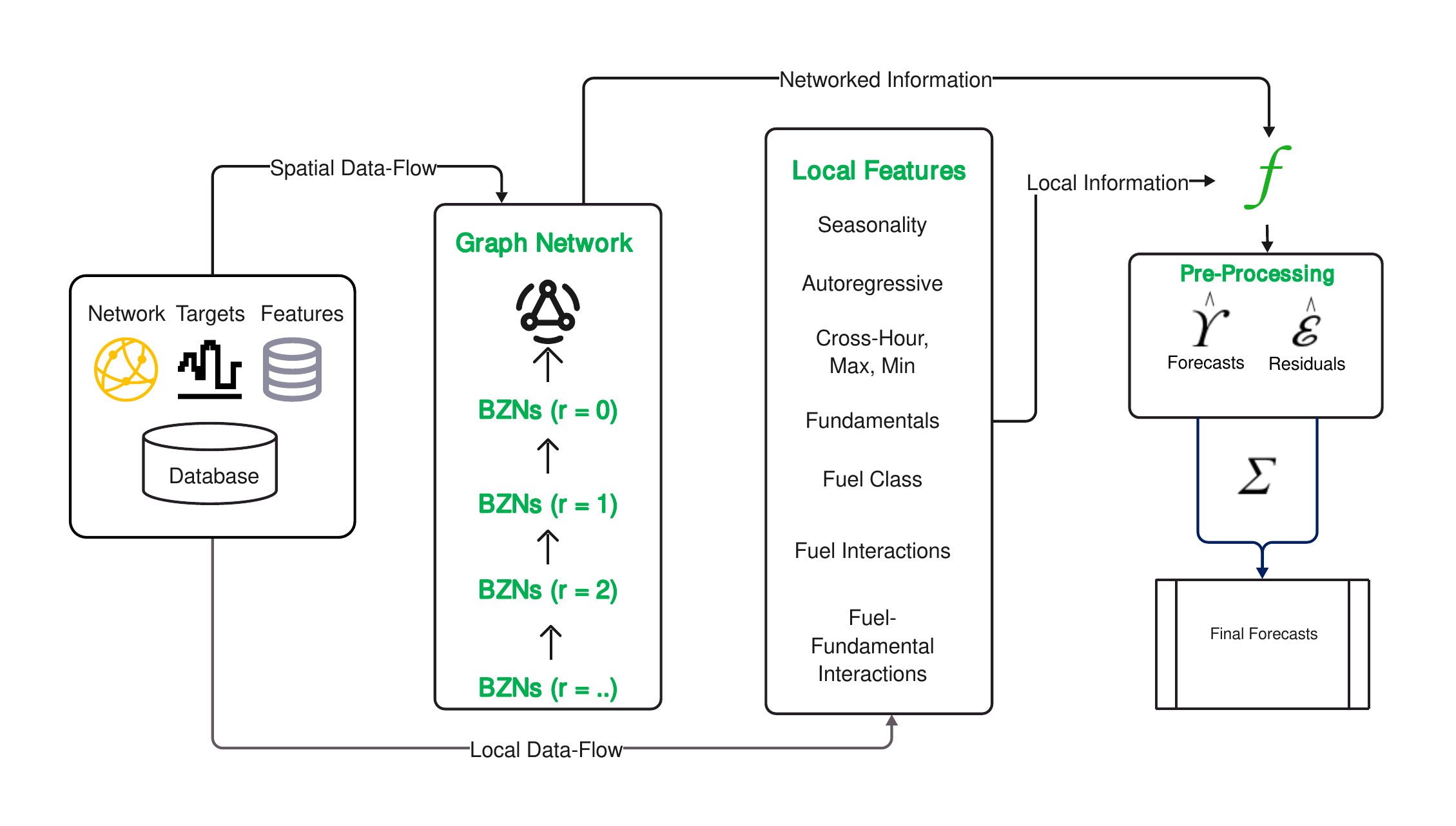}
    \caption{The Networked Spatio-Temporal Model's (NSTM) workflow begins with feature, network, and target databases. Information within a graph network structure at different radii $r=0,1,2,\dots$ forms the basis of the spatial spillover effects, while local features include seasonality, autoregressive terms, cross-hour statistics, fuels, fundamentals, and interactions. We implement a streaming setup in which information flows daily, and models are updated sequentially. Forecasts are generated in parallel across all bidding zones using 24-hourly models. A pre-processing step produces forecasts $\hat{Y}$ and residuals $\hat{\varepsilon}$, which are aggregated by a post-processing model to obtain the final one-step-ahead forecasts for all 39 bidding zones.}
    \label{fig:Workflow}
\end{figure}

One of the most important model classes in electricity price forecasting has been the statistical models, which have evolved into a major benchmark. Among statistical models, the ARX (Autoregressive with exogenous variables) is quite popular, as it integrates time-series components with a wide range of external variables. The classical ARX models have been enriched with a high-dimensional setup with shrinkage methods and a range of seasonal and exogenous factors in the so-called expert model class (\citet{zielricksven2015},\citet{ziel2016}, \citet{bertoszunejewskiweron2016},\citet{NowotarskiWeron2016}, \citet{ZielWeron2018},\citet{rafalbartosz2018}). We propose a model class, the Networked Spatio‑Temporal Model (NSTM), that extends the expert‑model paradigm by introducing networked information flow across interconnected markets.

\begin{figure}[H]
  \centering
  \includegraphics[width=1\linewidth, height=0.45\textheight, keepaspectratio]
  {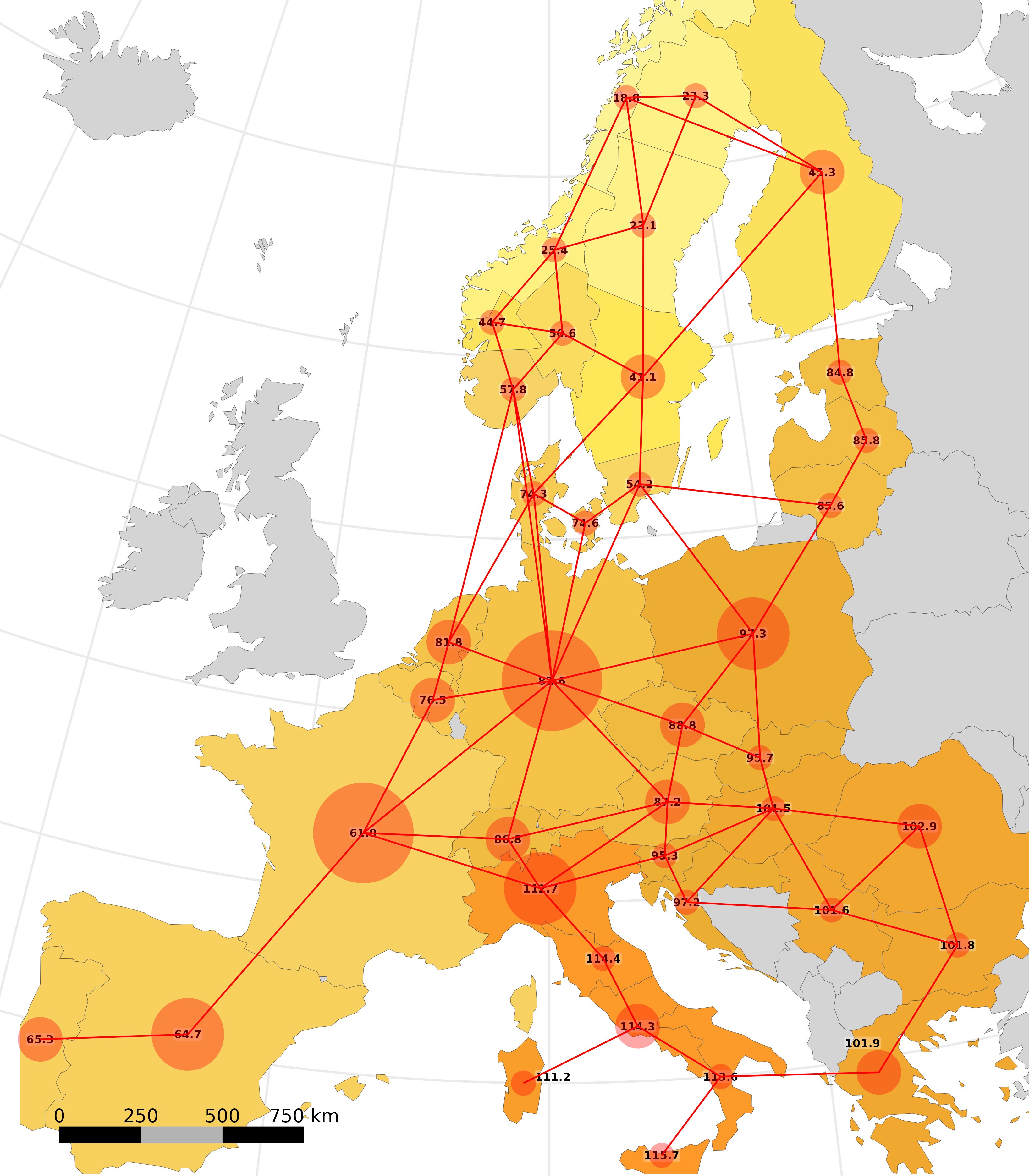}

  \captionsetup{
    justification=justified,
    singlelinecheck=false,
    font=footnotesize
  }

  \caption{
  European grid network with load and price characteristics. The figure illustrates the price and load profile of the bidding zones during the test period (Oct 2023-Sep 2025). The load is represented by circle size (larger circles $\rightarrow$ higher load),
  while prices are indicated by color intensity (darker color $\rightarrow$ higher prices).
  }

  \label{grid_graph}
\end{figure} 
We implement the NSTM across 39 bidding zones in the majority of European day‑ahead electricity markets in a high‑resolution, streaming-forecasting setting. The model captures key temporal regularities with autoregressive, cross‑hour, and seasonal effects, while incorporating fuel and emission prices and day‑ahead forecasts of fundamental variables as predictors within the network. Empirical results show that the proposed approach consistently outperforms traditional island‑based local models, both in zone‑specific accuracy and in aggregated, load‑weighted performance measures. \vspace{.2cm}

A central contribution of this work is the demonstration of systematic spatial effects through a rigorously specified spatial transmission mechanism. Incorporating these spatial effects in a networked structure yields superior forecasting performance relative to purely local models while preserving interpretability. The remaining paper is structured as follows. Section 2 describes the graph‑network structure underlying the information flow. Sections 3 and 4 present the data and the models. Section 5 details the estimation methodology, study design, evaluation protocol, and benchmarks. Section 6 reports the empirical results, and Section 7 concludes with final remarks and an outlook.

\section{Graph network} \label{sec:graph} The structure of spatial interactions in an integrated electricity market is difficult to capture with a single measure. Researchers have applied correlation, regression coefficients, and causal relationships across variables such as prices, congestion, import-export, Net Transfer Capacity (NTC), Available Transfer Capacity (ATC), Commercially Scheduled flow (CS), Net flow, etc. For details, please see \cite{cb1_deblauwe2025}. Also, \cite{CBDo2024} used the Diebold-Yilmaz (DY) framework from \cite{DieboldYilmaz2008}, \cite{DieboldYilmaz2012}. Also the Quantile Connected measure by \cite{QC_Ando2022}, Directed Acyclic Graphs (DAGs) by \cite{CBPark2006} and apllied network graphs \cite{CBYang2024} are some noteworthy approaches. \vspace{.2cm}

We used a Europe-wide metric graph to capture spillover relationships and systematically map information flow for a scalable study design. We denote day ahead electricity price by the random variable $Y_{d,s,z}$ is an element of the collection $Y_{d,S,\mathcal{V}}$, where $\mathcal{V}$ is the set of all spatial elements and $s \in S = \{0,1,2,\ldots,23\}$ is the intraday hours of a certain day. The process $Y_{d,S,\mathcal{V}}$ is a spatio-temporal object with a spatial dimension within a graph structure, and a mixed-frequency temporal dimension comprising both daily and hourly indices. We define the network structure to capture spatial relationships by formalizing it as a graph $G = (\mathcal{V}, \mathcal{E})$. For further discussion of graphs, see \citet{Kolaczyk2020}. The $\mathcal{V}$ is the set of vertices (commonly called the nodes), and $\mathcal{E}$ is the set of edges (commonly called the links) connecting these vertices. We apply two explicit restrictions regarding the spatio-temporal process.

\begin{itemize}
  \item \textit{We define the network in terms of the existence of a physical electricity transmission grid connection. The network connection between two spatial neighbours will be established only when a physical electric transmission line connects the two grids and electricity is flowing. This forms the basis of a first-degree connection.}
  \item \textit{The spatial process is ordered inside the network, and we only deal with the network paths that have the least distance. Within the network, there may be multiple paths connecting the two spatial locations, but we will consider only the shortest path.} 
\end{itemize}

\noindent The spatial properties with varying degrees of spatial distances are mapped by the adjacency matrix \(\mathcal{A}^k\), which is a mapping of the $k$-th distance of a certain neighbor with respect to an assessed price zone. An entry \(\mathcal{A}^k_{i,j} = 1\) indicates the $i$ and $j$-th zones are connected by an edge of length \(k\) (where \(k\) is the distance between the two zones). We propse the $G$ is undirected, thus $\mathcal{A}$ is symmetric (that is, $\mathcal{A}_{ij} = \mathcal{A}_{ji}$). We define the distance $k$ as the function $d: \mathcal{V} \times \mathcal{V} \to \mathbb{N}_0$ on the graph $G$. For any two zones $z_i, z_j \in \mathcal{V}$, the function $d(z_i, z_j)$ is defined as, 

\begin{equation}
k = d(z_i, z_j) =
\begin{cases}
0, & \text{if } z_i = z_j, \\[4pt]
\text{shortest path distance from } z_i \text{ to } z_j, & \text{if } z_i \neq z_j .
\end{cases}
\label{eq:distance}
\end{equation}

\vspace{0.2cm}

\noindent In this paper, the distance $d(\cdot,\cdot)$ is nonnegative and integer-valued. Obviously, a node not connected to any others eventually becomes a purely local process. We define a neighborhood with radius $r$ centered around a vertex $z \in \mathcal{V}$ as the set of all vertices with all  $k$ distances such that $k \le r$ with respect to $z$. Formally, the neighbourhood elements of radius $r$ centered at $z$ is expressed as:
\begin{equation}
 B_z^r = \{\, x \in \mathcal{V} \mid d(z, x) \le r \,\}
    \label{eq:ball}
\end{equation} So the $B_z^0=\{z\}$ denotes the neighborhood ball containing only the vertex $z$. We do not assume a uniform-global spatial process, but rather a spatial process unique to each bidding zone, in which the network's strength can be localized within a defined distance and neighborhood. So, the localization is a phenomenon within a certain level of spatial coverage within the network. Then incrementally, the neighborhood ball notation $B_z^1$ with $r=1$ is the set of all vertices at distance 1 or lower. So, for an arbitrary price zone $z$, the neighborhood ball includes all vertices at $k \le r$ of $z$ distance, including vortex $z$ itself, i.e., $B_z^1=\{z\}\cup\{\text{all first-degree neighbors of } z\}$. Then, as the general expression, for any finite integer $r\ge 0$, the neighborhood ball $B_z^r$ can be constructed using the equation \ref{eq:ball}. We have considered 39 bidding zones to capture the spatial properties of interconnected European electricity markets. Although a particular network may be fragmented into multiple clusters, with no direct path connecting them, the European electricity market is highly connected, and all price zones eventually become connected to each other.  In this study, all 39 bidding zones are eventually connected by a radius $r \leq 11$ (Figure \ref{grid_graph}). It is worth noting that each bidding zone has a different number of neighbors at a given radius $r$ or an adjacent neighbors at a certain distance $k$. So, the pace at which a bidding zone interconnects with the other nodes as we increase $r$ varies across zones. We have used separate notations, for clarity, to denote the distance measure $k$ and the radius measure $r$ for the neighborhood specification. As the mapping suggests, the radius $r$ is strictly nonnegative, and the neighborhood elements are monotonically increasing with $r$. So the adjacency matrix and neighborhood ball are strictly BZN-specific, and each BZN will have its own grid dynamics. The sets of BZNs at different distances $k$ (degree of adjacency) from a bidding zone are mutually exclusive. As the neighborhood radius maps the cumulative set of BZNs with adjacencies fulfilling the condition $k \leq r$, the element sets with neighborhood balls with increasing $r$ are a monotonically increasing set of elements. On the other hand, given a neighborhood ball mapping for a certain bidding zone $z$, the adjacency BZN set with $\mathcal{A}_z^k$ can be derived as follows,

\begin{equation}
  \mathcal{A}^k_z = B_z^{r=k} \setminus B_z^{r=k-1}
  \label{adjacency}
\end{equation}  \noindent So, the adjacency matrix $\mathcal{A}^k_z$ will denote the set of vertices whose distance from $z$ is exactly $k$. The neighboourhood ball thus can be mapped into a single spatial contigual matrix with dimension $|\mathcal{V}| \times |\mathcal{V}|$ where the elements are binary valued as well as each of the row elements maps all the bidding zones according to a specified radius $r$, if the zone is inside or on the radius $r$ the value 1 assigned otherwise zero. In the same way, the adjacencies can be mapped: adjacent zones with a particular distance $k$ have a value of 1, and the rest are zero.

\section{Data} This forecasting study includes data from 39 European bidding zones. The data for the empirical study are collected from the ENTSOE (European Network of Transmission System Operators for Electricity) transparency platform \citet{entsoe} and the European Energy Exchange \citet{eex}. We cover a considerable number of European price zones that encompass all major European countries, including the Nordics (Denmark: DK1, DK2; Sweden: SE1–SE4; Norway: NO1–NO5; and Finland: FI), the Baltic states (Latvia: LV, Lithuania: LT, and Estonia: EE), and part of the Balkan States (Croatia: HR, Serbia: RS, and Slovenia: SI), the Iberian Peninsula (Spain: ES and Portugal: PT), South-Eastern Europe (Romania: RO, Bulgaria: BG and Greece: GR), Central European states (Germany–Luxembourg: DE-LU, France: FR, The Netherlands: NL, Belgium: BE, Poland: PL, Croatia: HR, Czech Republic: CZ and Austria: AT) and Italy (IT-NORD, IT-CNOR, IT-CSUD, IT-SUD, IT-SARD, IT-SICI, IT-CALA). In total, we have covered 39 European bidding zones. We have not included the price zones from parts of the Balkan states, such as Albania (AL), Kosovo (XK), Montenegro (ME), and North Macedonia (MK), due to data issues. We also do not cover Great Britain (GB), Ireland (IE), Ukraine (UA), Belarus (BY), or Russia-controlled Kaliningrad. We build a spatial model variant based on the interconnected properties of European power grids, represented as spatial graphs as described in the graph network section \ref{sec:graph}. Due to the changing nature of European power grids, capacity extensions and new connections, we need to be aware that the salient characteristics of the grid are also evolving, or at least capacities are fluid over the assessed time. We set the network state as per the last availabe trainig date as a compromise to simplify the model somewhat. Also, as the United Kingdom and Ireland has substantial renewable generation and demand, not considering these two regions will inevitably affect the spatial models of the nearby bidding zones as well. Significant network developments during the sample period include the split of Austria (AT) from the Germany–Luxembourg price zone (DE-LU) and the split of the Italy Calabria zone (IT-CALA) from the Italy South zone (IT-SUD). As of the cutoff date for the network state fixation after Austria's split from Germany and Luxembourg, we adjust the connections of specific Eastern European vertices to the DE-LU zone and introduce IT-CALA as a new zone in a major update. All networks are constructed based on inter–price zone electricity export and import data. All forecasting of the respective price zones is done in CET, and the time is duly adjusted from the ENTSO-E reported UTC zone, including Daylight Saving Time (DST) adjustments. The data generally can be divided into the following 5 subgroups, and the respective particulars are as follows,

\begin{enumerate}
  \item \textbf{Target data:} Target data includes day-ahead electricity prices for all bidding zones. The data is collected at an hourly frequency. The primary data collected are denominated in EUR but, for different time periods, are also in local currencies such as Polish złoty (PLN), Romanian leu (RON), and Bulgarian lev (BGN). All prices are converted to Euro (EUR) using the relevant exchange rate at the time. Basis risk due to currency exchange is disregarded. Day-ahead prices are sourced from the ENTSO-E Transparency Platform, and exchange rate data is sourced from the EEX Data Platform.

  \item \textbf{Date-Time:} The corresponding timestamps are collected from the ENTSO-E in UTC format. The UTC format is then synchronised with Daylight saving adjustments and transformed into CET/CEST, depending on the datapoint, to properly adjust the auction calendar and local behavioural factors. These DST-adjusted timestamps serve as the basis for all seasonality adjustments and for forecasting operations' reference dates. Primarily, the date time is at the maximum quarter-hourly frequency to synchronise with some of the quarter-hourly data points, which are later uniformly processed to the hourly frequency for the forecasting data pipeline.

  \item \textbf{Fundamental data:} Fundamental data includes day-ahead forecasts for load, solar, and both onshore and offshore wind generation forecasts. These are collected at both hourly and quarter-hourly frequencies, depending on reporting changes. In some cases, simple transformations such as averaging are applied. After pre-processing, all fundamental data is synchronized to an hourly frequency.

  \item \textbf{Fuel prices:} Four fuel prices are used; the prices indicates the daily closing prices of monthly futues for coal, gas, EUA carbon price (EU ETS price), and oil. All prices are denominated in EUR currency. Coal and EU ETS prices are in tonnes; gas is in per megawatt hour (MWh); and oil is in barrel. Prices are collected daily at the end of each day.  For this study, we use T-2 day-end prices due to the auction schedule and the availability of pricing data. On any given day, all hourly models use the same fuel prices. Data is collected from the European Energy Exchange (EEX) Data Platform.

  \item \textbf{Network data:} Network data is constructed from export and import relationships across all 39 bidding zones. Details on network construction are discussed in the Graph Network section. Export-import data is collected from ENTSO-E. The reference time is the last day of the training data.

\end{enumerate}

\begin{figure}[ht!]
  \captionsetup{justification=justified,singlelinecheck=false, font= footnotesize}
  \includegraphics[width=1.0\textwidth]{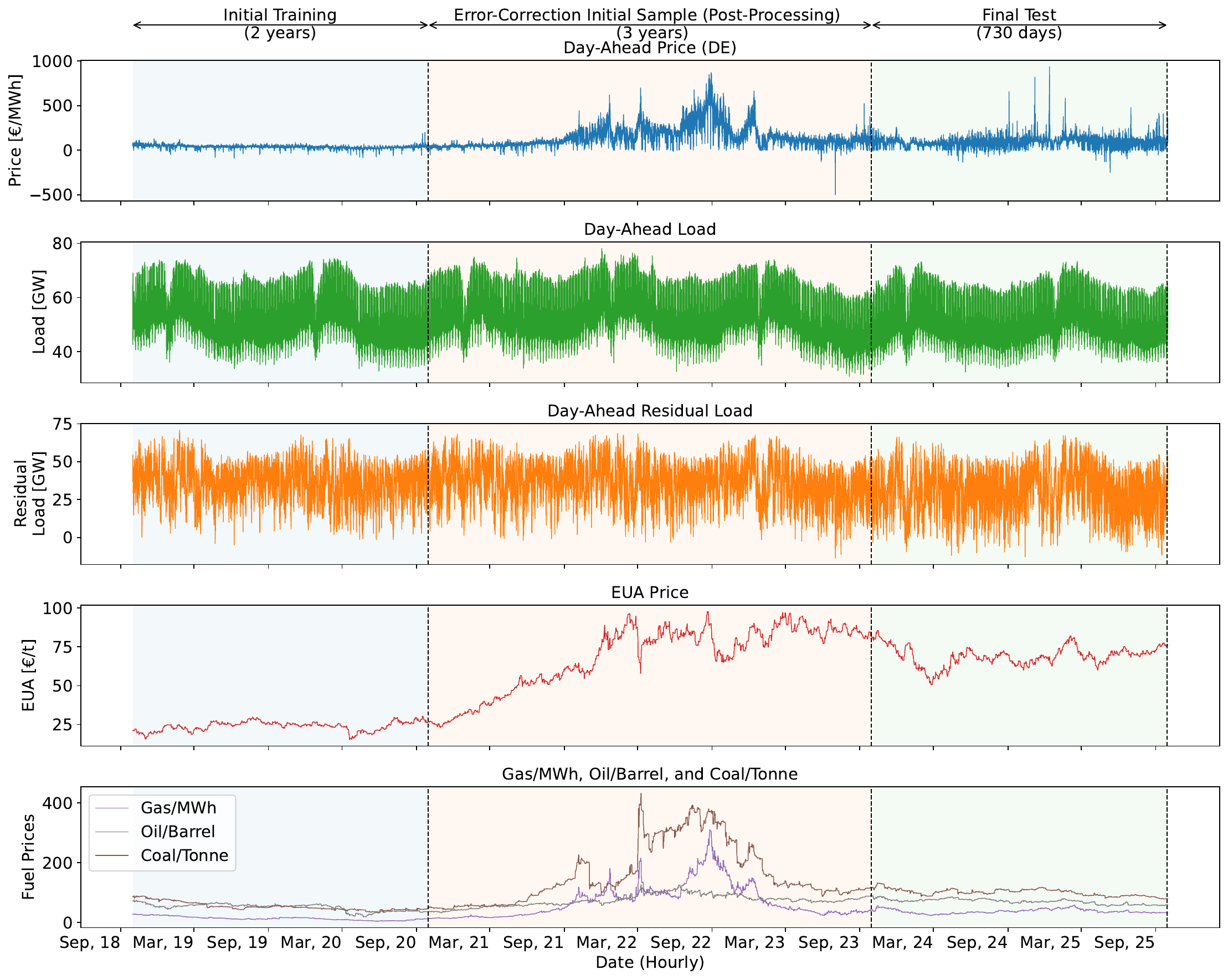}
  \caption{The figure above shows the data and study timeline for the German bidding zone as an example of standard setting. The study uses a two-stage approach: an initial 2-year training period, followed by 1-step day-ahead online forecasting for 5 years. The estimated residuals are then used in a 3-year rolling window for error correction, resulting in the final forecast for the last 2 years. }
  \label{study_design}
\end{figure}

We maintain certain strict characteristics throughout this paper, such as a synchronous data pipeline. Although there are sources of available information for some data points, or asynchronous market data (\cite{zielricksven2015} used EXAA market data, and \cite{cb_mascarenhas2026} used both Switzerland and EXAA market data, where the cross-market available auction prices are used), as we are forecasting a large number of bidding zones, we have not used asynchronous market information. All the timescales are strictly synchronized with CET/CEST, and the data is processed in daily updates. The stylized data are presented in Figure \ref{study_design}, with the German bidding zone as an example. The data is pre-processed to handle missing values and outliers and ensure consistency across bidding zones. The data is then used to train online streaming models, with the first pre-processing forecast spanning October 2020 to September 2025 (5 years). Then, in the post-processing stage, we apply error correction using a 3-year rolling window to arrive at the final forecast for the test sample, which runs from October 2023 to September 2025 (2 years). The seasonality is evident in the load and residual load data, whereas the fuel class shows an energy shock during 2021-2023. The volatility of prices also varies during the training and testing periods, along with changing levels of renewables integration, which plausibly influence the special relationship over time. To capture the changing nature of the regimes and seasonality, the data and modeling approach uses a streaming setup and strictly synchronous data to foster comparability within a nested modeling architecture.

\section{Models} 
\subsection{Modeling approches and related challenges}
The number of bidding zones in the European electricity market is substantial, and each has its own stylized behavior that significantly influences the DA-price equilibrium. So the BZN-specific modeling, either with hourly models or a unified model for all hours, is quite a reasonable direction. To an extreme, instead of BZN-specific modeling, we could model the whole of Europe with a single large model, including all the Bidding Zones, within a global modeling paradigm. Each approach has its pros and cons, with different motivations, but we must not forget that there are many heterogeneous patterns to negotiate in incredibly complex price processes. For instance, the intraday and across-the-day relationships differ significantly over time, creating a temporal effect with multiple seasonality signatures at different frequencies. In terms of spatial characteristics, treating each bidding zone as a separate node results in an irregular spatial grid. Now, when defining a joint spatio-temporal process with a clustered geographical pattern, the model specification may not be adequately handled by a singular monolithic model; rather, by a flexible, well-defined architecture that allows these idiosyncrasies to be modeled within a complex spatio-temporal framework seems more reasonable. \vspace{.2cm}   

One approach in econometric modeling is to assume a global process or, at the other extreme, a local, island-based approach. There are classes of spatial panel data models suitable for Europe-wide modeling, and, at their core, the intricacy of spatial relationships is captured by the spatial weight matrix. On the other hand, when we assume the spatial process is highly localized, we model each bidding zone as an island node. If, for the time being, we assume the process to be global in nature, such that all interconnected European price zones are part of a single integrated process, a starting point may be the General Nesting Spatial model by \citet{SLX},
\[
\mathbf{Y} = \rho \mathbf{W}\mathbf{Y} + \alpha \boldsymbol{\iota}_{N_z} + \mathbf{X}\boldsymbol{\beta} + \mathbf{W}\mathbf{X}\boldsymbol{\theta} + \lambda \mathbf{W}\mathbf{u} + \boldsymbol{\varepsilon}.
\]
Here, the target price vector \(\mathbf{Y}\) contains single-step observed prices for \(N_z\) bidding zones, where \(\mathcal{V}\) denotes the set of all price zones. The matrix \(\mathbf{W}\) is the spatial weight matrix of dimension \(N_z \times N_z\). The scalar parameters are \(\rho\), \(\alpha\), and \(\lambda\), while \(\boldsymbol{\beta}\) and \(\boldsymbol{\theta}\) are parameter vectors associated with exogenous explanatory variables and exogenous interaction effects, respectively. This broad model structure nests a range of popular spatial models, such as the Spatial Lag of Exogenous Variables Model (SLX), the Spatial Durbin Model (SDM), the Spatial Autoregressive Combined Model (SAC), the Spatial Error Durbin Model (SEDM), the Spatial Autoregressive Model (SAR), and the Spatial Error Model (SEM). Another popular approach for modeling spatially non-stationary processes was proposed by \citet{Brunsdon1998} using geographically weighted regression (GWR), expressed as \( \mathbf{Y}_z = \sum_j \mathbf{X}_{zj}\beta_j + \boldsymbol{\epsilon}_z \). The regression coefficients \( \hat{\boldsymbol{\beta}} \) can be estimated via weighted regression as \( \hat{\boldsymbol{\beta}} = (\mathbf{X}'\mathbf{W}_z\mathbf{X})^{-1}\mathbf{X}'\mathbf{W}_z\mathbf{y} \), where both the coefficient estimates and the diagonal weight matrix \( \mathbf{W}_z \), are zone-specific. As already discussed above, one of the most important aspects of spatial models is that they generally depend on a spatial weight matrix to represent spatial effects. Defining such a matrix is subject to notable debate and difficulty \citet{SLX}. For a highly defined process, a global weight matrix may be appropriate; however, for heterogeneous electricity price processes, specifying appropriate weights is very difficult. In a data-driven approach where the weights are not predefined but estimated, the complex nature of the weights may not be optimally specified.  For a non-homogeneous cluster, it may even be inconsistent. (In case the process is spatially non-stationary). Even if the weight matrix can be estimated, strong local influence, along with non-linear effects with spatial components, may complicate the estimation. In this backdrop, a valid question is which modeling framework is appropriate: a local, pure-play approach or a wide-scale, global model integration. The reason the localized forecasting models have been highly successful is that they can reasonably approximate the generalized dynamics. However, due to the integrated nature of the European electrical grids, neighboring bidding zones exert significant influence, and this influence is expected to increase in the future. 

\begin{figure}[H]
\centering
\captionsetup{justification=justified,singlelinecheck=false,font=footnotesize}
\includegraphics[width=.95\textwidth]{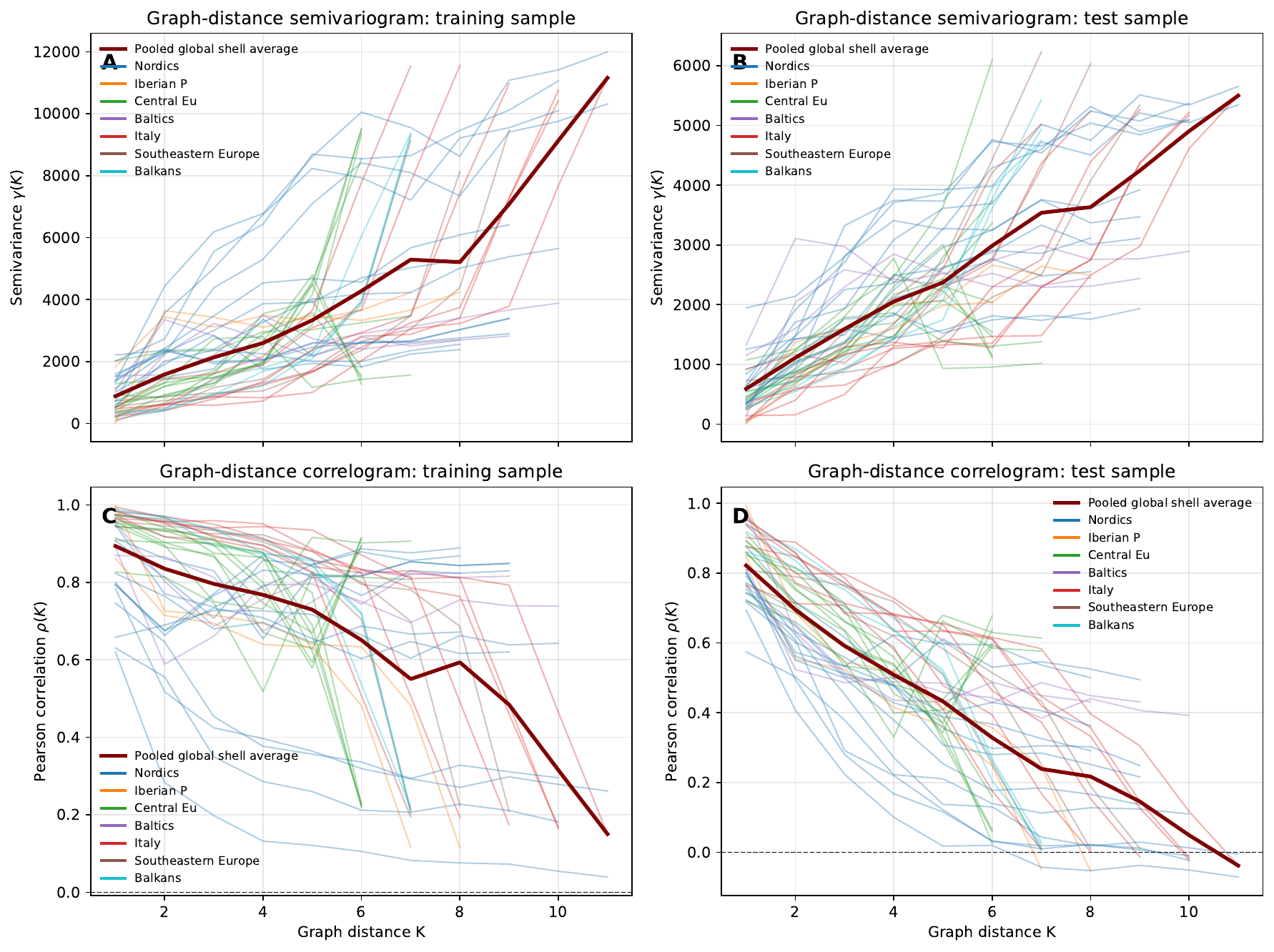}
\caption{
Panels A--B show the graph-distance semivariogram for the training and test samples, respectively, where each value is one half of the mean squared price difference for zone pairs at exact graph distance \(K\). Panels C--D show the corresponding graph-distance correlogram, defined as the mean Pearson correlation of zone-pair price series within each graph-distance shell.
}
\label{semivar_corr}
\captionsetup{font=footnotesize}
\end{figure}

Spatial dependence in the Figure \ref{semivar_corr} is evaluated using adjacency-distance classes on the transmission network $G=(\mathcal{V},\mathcal{E})$ defined in Section~\ref{sec:graph}. The horizontal axis represents adjacency distance $k$, where zones in the shell $\mathcal{A}_z^k$ satisfy $d(z_i,z_j)=k$ according to (\ref{eq:distance})--(\ref{adjacency}). The empirical semivariogram is computed as the average half-squared price difference across all ordered zone pairs within each adjacency shell, while the semicorrelogram reports the corresponding average pairwise correlation. The darker curve shows the global statistic averaged across all zones, and the lighter curves show zone-level estimates. The results indicate that spatial dependence in electricity prices follows the structure of the transmission network. Stronger co-movement is observed among zones with smaller adjacency distances, while dependence gradually declines as the distance increases. This suggests that expanding adjacency sets in the forecasting model may improve predictive performance by capturing network-based spillovers. However, the correlation structure tends to change during seasons and also over time. Also, although the average curve appears somewhat monotonic, it has kinks, and its shape is not stable over time. On the other hand, the individual bidding zones exhibit quite heterogeneous curves, and geographic clustering is observed across various segments of Europe. So instead of explicitly defining a weight matrix, we aim to systematically and flexibly expand the neighbourhood sets to dynamically capture spatial network effects for each bidding zone using the proposed NSTM. \vspace{.2cm}

So we take the view that neither the pure local nor the fully global approach alone would satisfy, for the following key reasons, (1.) The electricity price auction  mechanism starts within a defined bidding zone, where local supply and demand influence the price process by dominating the order book. However, in the presence of cross-border electricity flows, neighboring zones also become parties to the process, fuzzing the boundary of localization. The Pan-European Hybrid Electricity Market Integration Algorithm (PCR-EUPHEMIA) goes beyond bidding zones and introduces constrained optimization. The price process is spatially non-stationary, so using a global weight matrix in an irregularly spaced spatial grid may be misleading. In the presence of nonlinear demand and generation profiles, transmission bottlenecks, and availability and capacity thresholds, spillover effects can be asymmetric  and even nonlinear. (2.) The time-series structure of electricity prices displays clear intraday and across-day relationships. So, a mixed-frequency temporal process needs to be implemented within a rather flexible framework that can capture such time-varying dynamics, possibly with a large number of intricate spatial interactions.

\subsection{Networked Spatio‑Temporal Model} 
The European electricity price auction process is subject to an EU-wide matching algorithm and network-constrained optimization. The market participants for each bidding zone submit orders (bids and offers) within the gate closure at 12:00 CET, and the prices are published simultaneously for all 24 hours for the next day, taking into account network constraints, capacities, order books, and social welfare. Renewable output, load, generation, and demand patterns across bidding zones vary over time during the day and throughout the year. To capture these time-varying effects, researchers often use separate hourly models in energy price forecasting (EPF). Although European electrical grids are highly interconnected, they often have bottlenecks and transmission limitations across cross-border grids. So, the transmission is not unconstrained for various reasons, including investment requirements, neighboring bidding zones' load capacity, or other arbitrary reasons. Due to constraints and the spatial location of a particular bidding zone, the spillover effect may also differ zone to zone over time and distance. As we already have indicated, this type of heterogeneity makes it especially difficult to define a consistent weight matrix. Also, we observe seasonal price dependencies, alongside weekly (weekdays vs weekends) and intraday seasonalities (peak vs off-peak hour prices), etc. On the other hand, as price-discovery mechanisms are coupled, intraday-specific factors propagate through the network, subject to contingencies and constraints along the way, to arrive at intraday price curves. In that backdrop, we take a high-resolution, high-dimensional, networked modeling approach. So we implement models at hourly resolution in a networked-spatial setup, consequently expanding the information set across European network nodes to include relatively high-dimensional data. In this modeling architecture, we map the interconnected bidding zones at varying distances to estimate spillover effects at hourly resolution using an accurately calibrated mean model, and capture systemic intraday effects using a post-processing extension. This approach enables us to capture the day-varying long-term mean trajectory and short-term intraday divergences.  We propose the networked spatio-temporal model (NSTP) as per equation \ref{eq:ntp}, where the mean process is defined by both pure local regressors and spillover effects from neighboring zones within a defined radius \(r\),
\begin{equation}
y_{d,s,z}
=
m_{z,s}\!\bigl(\mathbf X_{d,s,z}\bigr)
+
f_{z,s}\!\bigl(\mathbf W_{d,s,z},\boldsymbol\Xi\bigr)
+
\eta_{d,s,z}.
\label{eq:ntp}
\end{equation}
\noindent The first component captures systematic variation explained by observable regressors in the mean function ($m_{z,s}(.)$), while the second component represents latent spatio-temporal effects through a post-processing function $f_{z,s}(\cdot)$. The mean component may be expanded as,
\begin{equation}
m_{z,s}\!\bigl(\mathbf X_{d,s,z}\bigr)
=
\mathbf x_{d,s,z}^{\top}\boldsymbol\beta_{z,s}
+
g(\mathbf X,\mathcal F,z,r)\,
\widetilde{\boldsymbol\beta}_{z,s}^{(r)} .
\label{eq:meanmodel}
\end{equation}

\noindent Where $\mathbf x_{d,s,z}$ denotes the local regressors \ref{lreg} design and $\widetilde{\mathbf x}_{d,s,z}^{(r)} = g(\mathbf X,\mathcal F,z,r)$ denotes the network regressor \ref{netreg} design. The local regressor matrix contains predictors constructed exclusively from information observed in the assessed bidding zone $z$. Formally,
\begin{equation}
\mathbf x_{d,s,z}
=
\left(
1,\,
\mathbf Y_{d,s,z}^{\mathrm{AR}\top},\,
\mathbf DoW_{d}^{\top},\,
\mathbf X_{d,s,z}^{\mathcal F\top},\,
\mathbf X_{d-2}^{\mathcal C\top},\,
\mathbf X_{d,s,z}^{\mathcal C\times\mathcal F\top},\,
\mathbf S_{d}^{\top},\,
\mathbf X_{d-2}^{\mathcal I_{\mathrm{fuel}}\top},\,
\mathbf X_{d,s,z}^{\mathrm{sq}\top}
\right)
\label{lreg}
\end{equation}

\noindent Where $1$ denotes the intercept, $\mathbf Y_{d,s,z}^{\mathrm{AR}}$ collects autoregressive and cross-hour price regressors, $\mathbf DoW_{d}$ represents calendar effects such as day-of-week dummies, $\mathbf X_{d,s,z}^{\mathcal F}$ contains zone-specific fundamentals (contemporaneous and lagged), $\mathbf X_{d-2}^{\mathcal C}$ includes fuel and EUA-ETS prices lagged by two days, $\mathbf X_{d,s,z}^{\mathcal C\times\mathcal F}$ denotes interactions between fuel prices and local fundamentals, $\mathbf S_{d}$ represents deterministic seasonal components captured through Fourier terms, $\mathbf X_{d-2}^{\mathcal I_{\mathrm{fuel}}}$ contains pairwise fuel interaction terms, and $\mathbf X_{d,s,z}^{\mathrm{sq}}$ collects quadratic transformations of selected regressors. These predictors correspond to widely used expert-type regressors in electricity price forecasting models. Electricity prices may also be affected by economic fundamentals observed in neighboring bidding zones, driven by cross-border electricity flows and transmission constraints. In the empirical application, the fundamental information set is restricted to day-ahead residual load forecasts (contemporaneous day-ahead residual load and its one-day lag). Although there are other sources of renewable generation sources other than solar and wind, the residual load is simply the difference between the day-ahead load forecast and the day-ahead renewable generation forecast (Solar and Wind only), and it is a widely used variable in the EPF literature to capture the net effect of renewables on price formation. Formally,
\begin{equation}
\mathcal{F} =
\left\{
\mathrm{RL}^{\mathrm{DA}}_{d,s},
\mathrm{RL}^{\mathrm{DA}}_{d-1,s}
\right\}
\label{fund}
\end{equation}

To capture the spatial spillovers, network regressors are constructed using the neighbourhood ball $B_z^r$ defined earlier. The transformation $g(\mathbf X,\mathcal F,z,r)$ maps the global information universe $\mathbf X$ into a zone-specific spatial regressors by selecting only the predictors specified in the fundamental information set $\mathcal F$ (Equation \ref{fund}) from neighbouring zones within the neighbourhood ball. Formally,
\begin{equation}
g(\mathbf X,\mathcal F,z,r)
=
\widetilde{\mathbf x}_{d,s,z}^{(r)}
=
\left(
X^{j,f}_{d,s}
\right)_{\substack{j\in B_z^r \setminus \{z\} \\ f\in\mathcal F}}
\label{netreg}
\end{equation}

\noindent where $X^{j,f}_{d,s}$ denotes the regressors associated with fundamental variable $f$ observed in neighbouring zone $j$. Thus, the network regressor vector collects selected fundamentals from neighbouring bidding zones inside the spatial neighbourhood $B_z^r$, while restricting the included regressors to those specified in the set $\mathcal F$. Because the number of neighbouring zones varies across bidding zones and increases with $r$, the dimension of $\widetilde{\mathbf x}_{d,s,z}^{(r)}$ is zone-specific and increases monotonically with the expansion of the neighbourhood radius. The full predictor used in the mean model thus can therefore be partitioned as,
\begin{equation}
\mathbf X_{d,s,z}
=
\left(
\mathbf x_{d,s,z}^{\top},
\widetilde{\mathbf x}_{d,s,z}^{(r)\top}
\right)^{\top}
\end{equation}

\noindent Consequently, the term, $\sum_{j\in B_z^r}\sum_{f\in\mathcal F} \beta_{z,s}^{j,f} X_{d,s}^{j,f}$ used in the final model specificnation in equation~\ref{eq:spatial_model} representing the empirical expansion with networked predictors $g(\mathbf X,\mathcal F,z,r)$. This essentially creates spatially nested models with varying levels of regionality (controlled by the neighborhood radius $r$). After implementing the mean model, the residuals are decomposed into a latent spatio-temporal component and an idiosyncratic noise component as follows,
\begin{equation}
\varepsilon_{d,s,z}
=
f_{z,s}\!\bigl(\mathbf W_{d,s,z},\boldsymbol\Xi\bigr)
+
\eta_{d,s,z}
\label{eq:epsilon-decomp-synced}
\end{equation}
\noindent It is worth noting that although we are implementing separate hourly models, because the models share the same daily states, the residuals of the daily model may contain useful structures. Because the errors after implementing the mean model are unobservable, we use the model's residuals from first stage regressions to capture the latent states. We assume that the latent spatial field \(\boldsymbol\Xi_{d,z}\) satisfies \(\mathbb{E}[\boldsymbol\Xi_{d,z}\mid \boldsymbol X] = \mathbf 0\), which is consistent with the elementwise condition \(\mathbb{E}[\widehat{\varepsilon}^{(r)}_{d,s,z}\mid \textbf{X}] = \mathbf 0\).  Specifically, we implement equation~\ref{eq:f-approx-cok1} for the post-processing stage and approximate \(f_{z,s}(\mathbf W_{d,s,z},\boldsymbol\Xi)\) using lagged cross-hour residuals. Although for a specific hour we have 23 available intraday cross-hour day-lagged residuals, we focus on (+/-) 1 neighboring hours and use a week's lag ensemble to capture short-term seasonal effects.
\begin{equation}
\label{eq:f-approx-cok1}
  f_{z,s}\!\bigl(\mathbf W_{d,s,z},\,\boldsymbol\Xi\bigr)
  \;\approx\;
  \frac{1}{|L|}
  \sum_{\ell\in L}
  \mathbf w^{(\ell)\!\top}_{d,s,z}\,
  \boldsymbol\Xi_{d-\ell,z}
\end{equation}

\noindent Finally the full specification of the Networked Spatio-Temporal Model (NSTM) specification with nested expert components is represented in the equation \ref{eq:spatial_model},

{\small
\begin{equation}
\begin{aligned}
Y_{z,d,s}
= \;& \beta_{z,s}^0
+ \underbrace{\sum_{p \in \PP} \beta_{z,s}^{\mathrm{auto},p}\,Y_{z,d-p,s}
+ \beta_{z,s}^{\mathrm{min}}\,Y_{d-1,\mathrm{min}}
+ \beta_{z,s}^{\mathrm{max}}\,Y_{d-1,\mathrm{max}}
+ \beta_{z,s}^{\mathrm{last}}\,Y_{z,d-1,S-1}}_{\text{Autoregressive \& cross-hour}} \\[2pt]
&+ \underbrace{\sum_{w \in W} \beta_{z,s}^{w}\,\mathrm{DoW}^{\,w}_d}_{\text{Calendar effects}}
+ \underbrace{\sum_{f\in \FF}\beta_{z,s}^{f}\,X^{{z,f}}_{d,s}}_{\text{Fundamentals}}
+ \underbrace{\sum_{c \in \CC} \beta_{z,s}^{c}\,X^{c}_{d-2}}_{\text{Fuel and EUA-ETS}}
+ \underbrace{\sum_{c\in \CC}\beta_{z,s}^{\,c\times \mathrm{RL}}\,X^{c}_{d-2}\,\mathrm{RL}^{\mathrm{DA}}_{z,d,s}}_{\text{Fuel}\times\text{residual load}} \\[2pt]
&+ \underbrace{\sum_{k=1}^{3}\!\Big(
\gamma_{z,s}^{\cos,k}\,\cos\!\frac{2\pi k\,\mathrm{DoY}_d}{365.25}
+ \gamma_{z,s}^{\sin,k}\,\sin\!\frac{2\pi k\,\mathrm{DoY}_d}{365.25}
\Big)}_{\text{Annual Fourier }(k=1,2,3)}
+ \underbrace{\sum_{(c_1,c_2)\in \mathcal I_{\mathrm{fuel}}}
\beta_{z,s}^{\,c_1\times c_2}\,X^{c_1}_{d-2}\,X^{c_2}_{d-2}}_{\text{Fuel interactions}}\\[2pt]
&+ \underbrace{\sum_{c \in \CC} \beta_{z,s}^{c^{sq}}\,\big(X^{c}_{d-2}\big)^{sq}
+ \beta_{z,s}^{\mathrm{RL}^{sq}}\,\big(\mathrm{RL}^{\mathrm{DA}}_{z,d,s}\big)^{sq}}_{\text{Quadratic}} + \underbrace{\sum_{j\in B_{z}^r}\sum_{f\in\mathcal{F}}
            \beta_{z,s}^{j,f}\,X_{d,s}^{\,j,f}}_{\text{Spatial effects}} \\[2pt]
&+\;\;\underbrace{\frac{1}{|L|}\sum_{\ell\in L}\mathbf w^{(\ell)\!\top}_{d,s,z}\boldsymbol\Xi_{d-\ell,\;z}}_{\text{Post-processing}}
\;+\; \eta_{z,d,s}\,.
\end{aligned}
\label{eq:spatial_model}
\end{equation}
}
\vspace{.2cm}

\noindent\textbf{The model components are:}
\begin{enumerate}
  \item \textbf{Autoregressive \& cross-hour:}
        Lags \(\PP=\{1,2,7\}\) of the same zone–hour,
        the previous-day extrema \(Y_{d-1,\min}\), \(Y_{d-1,\max}\),
        and the last hour of the previous day \(Y_{z,d-1,S-1}\).
  \item \textbf{Calendar effects (DoW):}
        Day-of-week dummies \(\{\mathrm{DoW}^w_d\}_{w\in W}\) to capture weekly patterns.
  \item \textbf{Fundamentals:}
        Zone-specific contemporaneous regressors \(X^{z,f}_{d,s}\) for \(f\in\FF\);
        this universe includes the DA residual load \(\mathrm{RL}^{\mathrm{DA}}_{z,d,s}\)
        and its lagged variant(s) (e.g., lag 1).
  \item \textbf{Fuel and EUA-ETS prices:}
        Fuel-class prices \(X^{c}_{d-2}\) for \(c\in\CC=\{\text{coal, gas, oil, carbon}\}\),
        lagged by two days (using day-end futures as proxies for marginal costs).
  \item \textbf{Fuel \(\times\) DA residual load:}
        Bilinear interactions \(X^{c}_{d-2}\cdot \mathrm{RL}^{\mathrm{DA}}_{z,d,s}\) for \(c\in\CC\).
  \item \textbf{Annual Fourier seasonality:}
        Sine–cosine harmonics for \(k=1,2,3\) using day-of-year \(\mathrm{DoY}_d\):
        \(\cos\!\frac{2\pi k\,\mathrm{DoY}_d}{365.25}\) and
        \(\sin\!\frac{2\pi k\,\mathrm{DoY}_d}{365.25}\).
  \item \textbf{Fuel class interactions:}
        Pairwise fuel cross-terms \(X^{c_1}_{d-2}X^{c_2}_{d-2}\) for
        \((c_1,c_2)\in\mathcal I_{\mathrm{fuel}}\).
  \item \textbf{Quadratic terms:}
        \((X^{c}_{d-2})^{sq}\) for \(c\in\CC\) and
        \((\mathrm{RL}^{\mathrm{DA}}_{z,d,s})^{sq}\).
  \item \textbf{Networked spatio-temporal effects:}
        Cross-regional fundamentals from neighbors
        \(j\in B_{z}^r\) (within radius \(r\)) at the same hour:
        \(X^{\,j,f}_{d,s}\) for \(f\in\FF\).
  \item \textbf{Latent Components:}
        Latent field \(\boldsymbol\Xi_{d-\ell,k}\) aggregated over
        lags \(\ell\in L\)  with weights
        \(\mathbf w^{(\ell)}_{d,s,z}\), normalized by \(|L|\).
\end{enumerate}

\section{Empirical Application} 
\subsection{Study design} The forecasting study undertaken in this paper is a point-forecasting study aimed at obtaining the day-ahead expected electricity prices across a large number of bidding zones in a synchronous manner. Here, we test our Networked Spatio-Temporal Models with varying degrees of spatial coverage. We have used both Ordinary Least Squares and Lasso-based estimators to test the pure local and spatial models within a neighborhood ball of radius 5. As we conduct the study across a large number of bidding zones spanning 39 BZNs, we employ an online learning algorithm to accommodate the streaming nature of the forecasts and improve computational efficiency. We conduct the study in two stages. In the first stage, using the full dataset spanning seven years (October 2018 -- September 2025), we obtain the first-stage forecasts. The initial two-year sample (October 2018 -- September 2020) is used to estimate the initial coefficient forecasts. Subsequently, the models for all BZNs are updated daily and produce 24-hour-ahead forecasts using the streaming algorithm. In the second stage, we use three years of residuals to implement the post-processing step, finally arriving at the final test sample (October 2023 -- September 2025). Thus, in the streaming configuration, at the beginning of the third year (October 2020), we obtain the first forecast and residual for the mean model, and in the sixth year (October 2023), we obtain the first post-processed NSTM forecasts and the corresponding residuals. The second stage is essentially a post-processing step that approximates the intraday latent effect using the estimated residuals from the first stage. \vspace{.2cm}

In the online setup, the coefficients are updated sequentially as new daily information becomes available, and forecasts for the next day are generated. We estimate 24 separate hourly models for each bidding zone (BZN). After generating forecasts for a given day, the procedure advances to the next day in a streaming manner until the final day's prices are forecast. We use both OLS and Lasso-based estimation methods, implement the NSTM. In the absence of any forgetting factor or observation weighting, the Online-OLS estimation procedure corresponds to an expanding-window setup, in which all previous observations are included in the objective function with equal importance. In our modeling framework, the final design matrix \(\mathbf X\) comprises a relevant predictor/feature universe that includes expert-type model predictors with a defined fundamental mix and information aggregated across zones. The scope of the spatial effects is controlled by the radius measure $r$, which essentially determines the number of spatial components in the model.  Later, we define the base model as the pure local model and use it as the benchmark model. It is convenient that the benchmark model is a variant of the spatial model with no assumed spatial effects in the nested NSTM specification. 

\subsection{Estimation} We have implemented the online learning framework for both OLS and Lasso-based models because it achieves fair predictive performance while offering superior computational efficiency and runtime. For OLS-based models, parameter estimation is carried out using the online version of Recursive Least Squares (RLS) \cite{RLSMaulins2005}. RLS provides an efficient calculational procedure for obtaining the exact expanding-window OLS estimator without repeatedly refitting the model from scratch. For Lasso-based estimation, the estimator is obtained using Online Coordinate Descent (OCD) \cite{OCDAngelosante}, warm-started from the previous estimate~$\hat{\boldsymbol{\beta}}_{d-1}^{\mathrm{lasso}}$. At time~$d$, the online OLS estimator minimizes the expanding-window squared-error loss, while for Lasso-based estimation, we consider the standard lasso loss function (\eqref{lossf_online}). A separate regularization parameter~$\lambda_{d,s,z}$ is selected for each estimation hour and each bidding zone. The regularization parameter is re-estimated at each time step using the Akaike Information Criterion (AIC), allowing the degree of shrinkage to adapt as the information set expands. During the training phase, both the target variable and the design matrix are normalized using the location and scale parameters computed from the training sample only. These normalization parameters are held fixed during the online update phase to avoid look-ahead bias.  We deliberately refrain from introducing additional hyperparameters. This design option aims to reduce estimation risk in the presence of regime changes or extreme price events and assures that changes in the estimated coefficients are driven mainly by incoming data. 

\begin{equation}
\mathcal{L}_d^{\mathrm{OLS}}(\boldsymbol{\beta})
=
\frac{1}{2d}
\sum_{i=1}^{d}
\left(y_i - \mathbf{x}_i^\top \boldsymbol{\beta}\right)^2 
\label{lossfols_online}
\end{equation}

\begin{equation}
\mathcal{L}_d^{\mathrm{lasso}}(\boldsymbol{\beta})
=
\frac{1}{2d}
\sum_{i=1}^{d}
\left(y_i - \mathbf{x}_i^\top \boldsymbol{\beta}\right)^2
+
\lambda_d
\sum_{j=1}^{p}
\left| \beta_j \right| 
\label{lossf_online}
\end{equation}

\noindent The explicit dependence on~$d$ reflects both the expanding data window and the re-selection of the regularization parameter over time. Rather than explicitly recomputing the loss at each update step, online estimation updates the sufficient statistics associated with the squared-error term. Upon arrival of a new observation~$(\mathbf{x}_d, y_d)$, the following recursive updates are applied,
\begin{equation}
\begin{aligned}
\mathbf{G}_d &= \mathbf{G}_{d-1} + \mathbf{x}_d \mathbf{x}_d^\top, \\
\mathbf{h}_d &= \mathbf{h}_{d-1} + \mathbf{x}_d y_d.
\end{aligned}
\label{online_updates}
\end{equation}

\noindent where,
\[
\mathbf G_d = \mathbf X_d^{\top}\mathbf X_d,
\qquad
\mathbf h_d = \mathbf X_d^{\top}\mathbf y_d
\]
denote the Gram matrix of the regressors and the response cross-product vector, respectively. Both quantities can be updated recursively as new observations $(\mathbf {x} _ {d}, y _ {d})$ arrive. The updates are algebraically equivalent to expanding the loss function from~$ d-1$ to~$ d$ and constitute the computational backbone of both OLS and Lasso estimation in the online setting.
\begin{equation}
\hat{\boldsymbol{\beta}}_{d}^{\mathrm{OLS}}
=
\mathbf G_{d}^{-1}\mathbf h_{d}
=
\hat{\boldsymbol{\beta}}_{d-1}^{\mathrm{OLS}}
+
\frac{
\mathbf G_{d-1}^{-1}\mathbf x_{d}
}{
1+\mathbf x_{d}^{\top}\mathbf G_{d-1}^{-1}\mathbf x_{d}
}
\left(
y_{d}
-
\mathbf x_{d}^{\top}
\hat{\boldsymbol{\beta}}_{d-1}^{\mathrm{OLS}}
\right).
\end{equation}

\noindent For the Lasso, no closed-form solution exists due to the non-differentiability of the $\ell_1$-penalty. Expanding the quadratic loss yields the equivalent formulation,
\begin{equation}
\label{eq:lasso_cd_objective}
\hat{\boldsymbol{\beta}}_{d}^{\mathrm{lasso}}
=
\arg\min_{\boldsymbol{\beta}}
\left\{
\frac{1}{2}\boldsymbol{\beta}^{\top}\mathbf G_d\boldsymbol{\beta}
-
\mathbf h_d^{\top}\boldsymbol{\beta}
+
\lambda_d \|\boldsymbol{\beta}\|_1
\right\}
\end{equation}
Given the loss objective in \eqref{eq:lasso_cd_objective}, it can be solved efficiently via cyclic coordinate descent with the soft thresolding. Here, $\hat{\boldsymbol{\beta}}_{d}^{(k)}$ denote the coefficient vector at iteration $k$. Updating the $j$-th coordinate while holding all other coefficients fixed reduces the problem to a one-dimensional penalized quadratic minimization. \citet{hirsch2024online} extend the coordinate-descent Lasso algorithme of \citet{OCDAngelosante} and \citet{mesnner2019online}, for online update setting which is essentially the $j$-th coefficient update derived by applying the sufficient statistics  \(\mathbf G_d\) and \(\mathbf h_d\) to partial rediuals finally arriveing to the equation \eqref{eq:lasso_cd_update},
\begin{equation}
\label{eq:lasso_cd_update}
\hat{\beta}_{d,j}
\leftarrow
\frac{
\mathcal S\!\left(
h_{d,j}
-
\mathbf G_{d,j,\bullet}\hat{\boldsymbol\beta}_d
+
G_{d,j,j}\hat\beta_{d,j},
\ \lambda_d
\right)
}{
G_{d,j,j}
}
\end{equation}
The \(\mathcal S(z,\gamma)=\operatorname{sign}(z)\max(|z|-\gamma,0)\) is the soft-thresholding operator. \(\mathbf G_{d,j,\bullet}\) denotes the $j$-th row of \(\mathbf G_d\). The objective function in \eqref{eq:lasso_cd_objective} is convex whenever the Gram matrix \(\mathbf {G} _ {d} \) is positive definite. Cyclic coordinate descent generates a sequence of iterates that monotonically decreases the objective function and therefore converges to the global minimizer. In the online setting, the algorithm is warm-started from the previous solution.  When the data-generating process is smooth, the coefficient vector at time \(d\) remains close to that at time \(d-1\), \citet{OCDAngelosante}. We employ the Python package \texttt{Online Distributional Learning (ondil)} to estimate the online Lasso coefficients. For further methodological details, see \citet{hirsch2024online}. The Ondil framework focuses primarily on distributional forecasts, but since we are interested only in point forecasts, we evaluate only the mean parameter \((\mu)\). 

In the empirical model in equation~\eqref{eq:spatial_model}, the network-related design matrix is defined as \(\widetilde{\mathbf X}_{d,s,z}^{(r)} = g(\mathbf X,\mathcal{F}_i,z,r)\), which is a subset of the full information set \(\mathbf X\) which denotes the universe of all available interconnected information containing all time \((d,s)\) and bidding-zone \(z \in \mathcal{V}\) as well as potentially engineered features. Using the online setup, we first estimate the pre-processing forecasts and the corresponding errors over the full forecasting window. It uses the zone-specific predictors and spatially relevant information set derived by the \(g(\cdot)\). In the first stage, we implement the NSTM using equation \ref{eq:meanmodel}.  The corresponding lagged residual stacks use lag \(\ell \in L\), constructed with cross-hour indices \(\tilde{s} \in \mathcal{O}_s^{\ast}\) for the zone under assessment \(z\), at hour \(s\) and design with radius \(r\) is defined in Equation~\ref{lagresidualstack} using the stage-1 residuals \eqref{stage1residual},
\begin{equation}
\label{stage1residual}
\widehat{\varepsilon}^{(r)}_{d,s,z}
=
Y_{d,s,z}
-
\mathbf x_{d,s,z}^{\top}\widehat{\boldsymbol\beta}_{z,s}
-
\widetilde{\mathbf X}_{d,s,z}^{(r)\top}
\widehat{\widetilde{\boldsymbol\beta}}^{(r)}_{z,s}
\end{equation}

\noindent

\begin{equation}
\label{lagresidualstack}
\boldsymbol q_{d-\ell,s,z}
=
\bigl[
\widehat{\varepsilon}^{(r)}_{d-\ell,s+\tilde s_1,z},\,
\widehat{\varepsilon}^{(r)}_{d-\ell,s+\tilde s_2,z},\,
\dots,\,
\widehat{\varepsilon}^{(r)}_{d-\ell,s+\tilde s_{|\mathcal O_s^\ast|},z}
\bigr]^{\!\top}
\end{equation}
and, on a \(\mathcal W\)-day rolling window (e.g.\ \(\mathcal W=1095\)), we form design-specific covariance structures and implement the post-processing error correction using the intraday hourly structure around the neighborhood (specified by the offsets) for the hourly models. In particular, we apply emperical variance-covariance structure to forecast the target-day residuals with estimated weights. At the end the final forecast is result of a naive ensamble (Using day lag 1 to 7 intraday structure) average of forecasted errors.

\[
\bar{\boldsymbol q}^{(\ell)}_{d,s,z}
=
\frac{1}{\mathcal W}
\sum_{u=d-\mathcal W}^{d-1}
\boldsymbol q_{u-\ell,s,z},
\qquad
\bar{\varepsilon}^{(r)}_{d,s,z}
=
\frac{1}{\mathcal W}
\sum_{u=d-\mathcal W}^{d-1}
\widehat{\varepsilon}^{(r)}_{u,s,z}
\]

\[
\boldsymbol\Sigma_{d,s,z}^{(\ell)}
=
\frac{1}{\mathcal W-1}
\sum_{u=d-\mathcal W}^{d-1}
\bigl(\boldsymbol q_{u-\ell,s,z}-\bar{\boldsymbol q}^{(\ell)}_{d,s,z}\bigr)
\bigl(\boldsymbol q_{u-\ell,s,z}-\bar{\boldsymbol q}^{(\ell)}_{d,s,z}\bigr)^{\!\top}
\in \mathbb R^{p\times p}
\]

\[
\boldsymbol c_{d,s,z}^{(\ell)}
=
\frac{1}{\mathcal W-1}
\sum_{u=d-\mathcal W}^{d-1}
\bigl(\widehat{\varepsilon}^{(r)}_{u,s,z}-\bar{\varepsilon}^{(r)}_{d,s,z}\bigr)
\bigl(\boldsymbol q_{u-\ell,s,z}-\bar{\boldsymbol q}^{(\ell)}_{d,s,z}\bigr)
\in \mathbb R^{p},
\qquad
p = |\mathcal O_s^\ast|
\]
\noindent

\noindent We assume a mean-zero error structure for individual hourly models, but since we use a rolling window with a finite sample in the second stage model, we normalize to obtain a more stable estimate. The cross-hour residual stacks are often highly correlated and possibly redundant, causing instability in estimation. Ridge regularization ensures invertibility, controls estimation variance, and yields more stable and robust residual corrections. The lag-specific weights are defined as,
\[
  \mathbf w_{d,s,z}^{(\ell)}
  =
  \bigl(\bm\Sigma_{d,s,z}^{(\ell)}+\lambda\,\bm I\bigr)^{-1}
  \bm c_{d,s,z}^{(\ell)},
  \qquad
  \lambda
  =
  \rho\,\frac{\operatorname{tr}(\bm\Sigma_{d,s,z}^{(\ell)})}{p},
  \quad
  \rho=10^{-4}
\]
where \(p = |\mathcal O_s^\ast|\) denotes the dimension of cross hours considered. The object \(\mathbf W_{d,s,z}\) denotes a lagged, hour-specific weight matrix that incorporates all coefficient vectors \(\{\mathbf {w}^{(\ell)}_{d,s,z}\}_{\ell\in L}\) corresponding to the lag ensemble. All covariance objects  \(\bm{\Sigma}_{d,s,z}^{(\ell)}\) are estimated on a rolling window of \(\mathcal W\). Here, \(f_{z,s}(\cdot)\) represents the \(s\)-th hour–specific component of a mapping. Because of potential redundancy in the hourly information captured across multiple lags, we ultimately compute an ensemble average of the forecasts obtained by the past intrady information in each lag.

\subsection{Benchmark pure local and Naive model} The proposed NSTM is a nesting model of increasing level of spatial integration on top of the pure local island model. For instance, if we assume a disconnected-island structure, the main model can be reduced to a pure local model, equivalent to a design having neighbourhood ball of radius zero. Thus, the reduced model can be expressed as,

\begin{equation}
  y_{d,s,z}
=
\underbrace{\mathbf x^{(0)\!\top}_{d,s,z}\,\boldsymbol\beta^{(0)}_{d,s,z}}_{\text{pure local}}
\;+\;
\varepsilon_{d,s,z}
  \label{eq:base}
\end{equation}

\noindent we also use the parameter free naive model defined in the equation \ref{eq:naive},

\begin{equation}
y_{d,s,z}
=
\begin{cases}
y_{d-7,s,z}, & \text{if } \hspace{.25cm}d \in $\{Sat, Sun, Mon\}$, \\[4pt]
y_{d-1,s,z}, & \text{otherwise},
\end{cases}
\;+\;
\varepsilon_{d,s,z}
\label{eq:naive}
\end{equation} \vspace{.2cm}

\noindent If we examine the errors of both models, it is clear that the error structures of the benchmark and naive models differ, as $\varepsilon$ is not equal to $\eta$ as in the final NSTM in equation \eqref{eq:spatial_model}. We consider the base NSTMs, including the local models, in their pre-processed basis. After error correction, we obtain the final NSTM model, which is post-processed in the second stage. This also enables us to examine the model's performance, both the network effects and the post-processing error-correction contributions.

\subsection{Evaluation metrics} The two most popular accuracy measures in point forecasting are the Mean Absolute Error (MAE) and the Root Mean Squared Error (RMSE). In addition, we employ the Diebold-Mariano (DM) test, \citet{dm1} to assess relative forecast performance across competing models. The Root Mean Squared Error (RMSE) is defined as,
\begin{equation}
\label{rmse}
\mathrm{RMSE_z}
=
\sqrt{
\frac{1}{D S}
\sum_{d=1}^{D}
\sum_{s=0}^{S-1}
\bigl(
Y_{d,s,z}
-
\widehat{Y}_{d,s,z}
\bigr)^2
}.
\end{equation}
A lower RMSE indicates better point forecast performance. RMSE is commonly used to evaluate point forecasts of the conditional mean. The Mean Absolute Error (MAE) is defined as,
\begin{equation}
\label{mae}
\mathrm{MAE_z}
=
\frac{1}{D S}
\sum_{d=1}^{D}
\sum_{s=0}^{S-1}
\left|
Y_{d,s,z}
-
\widehat{Y}_{d,s,z}
\right|.
\end{equation}
MAE provides a complementary measure that is less sensitive to large forecast errors than RMSE. In addition, we use the Diebold-Mariano (DM) test. let \(L(\cdot)\) denote a common loss function (e.g.\ squared error \(L(e)=e^{2}\) or absolute error \(L(e)=|e|\)) and loss differential is defined as \(\Delta_{A,B,d,s,z}=L(e_{A,d,s,z})-L(e_{B,d,s,z})\). As the sample mean of \(\Delta_{A,B,d,s,z}\) is asymptotically normally distributed, the null hypothesis \(H_{0}:\mathbb{E}(\Delta_{A,B,d,s,z})=0\) is rejected at a given significance level \(\alpha\) if the test statistic exceeds the corresponding critical value from the standard normal distribution, thus indicating which model in the comparison pair performs better.

\section{Empirical Results} We found strong evidence that networked spatial structure impacts the performance of European electricity price forecasting, with NSTM models outperforming pure local models across European markets. A pure local base benchmark model can be considered as a bare-bones NSTM-Base model with no additional nodes beyond the assessed bidding zone (Hence $r$=0). We take the OLS version of the pure local model as the benchmark for all BZN and Europewide average performances. As $r$ increases, we achieve greater regional coverage by including the residual load information (Day-ahead and lag-1 forecasts) for all zones within the neighborhood ball. NSTMs with $r>0$ are all spatial model variants defined by the various depths ($r$), essentially capturing increasing spillovers within the specified radius. We also incorporate the post-processing to arrive at the NSTM post-processed models, which is essentially our final model class. The NSTM model class with and without post-processing, substantially improves the model's predictive performance. On top of that, the post-processing correction provides an additional performance boost. 
\begin{figure}[H]
  \centering
  \captionsetup{justification=justified,singlelinecheck=false, font=footnotesize}
  \includegraphics[width=.95\textwidth]{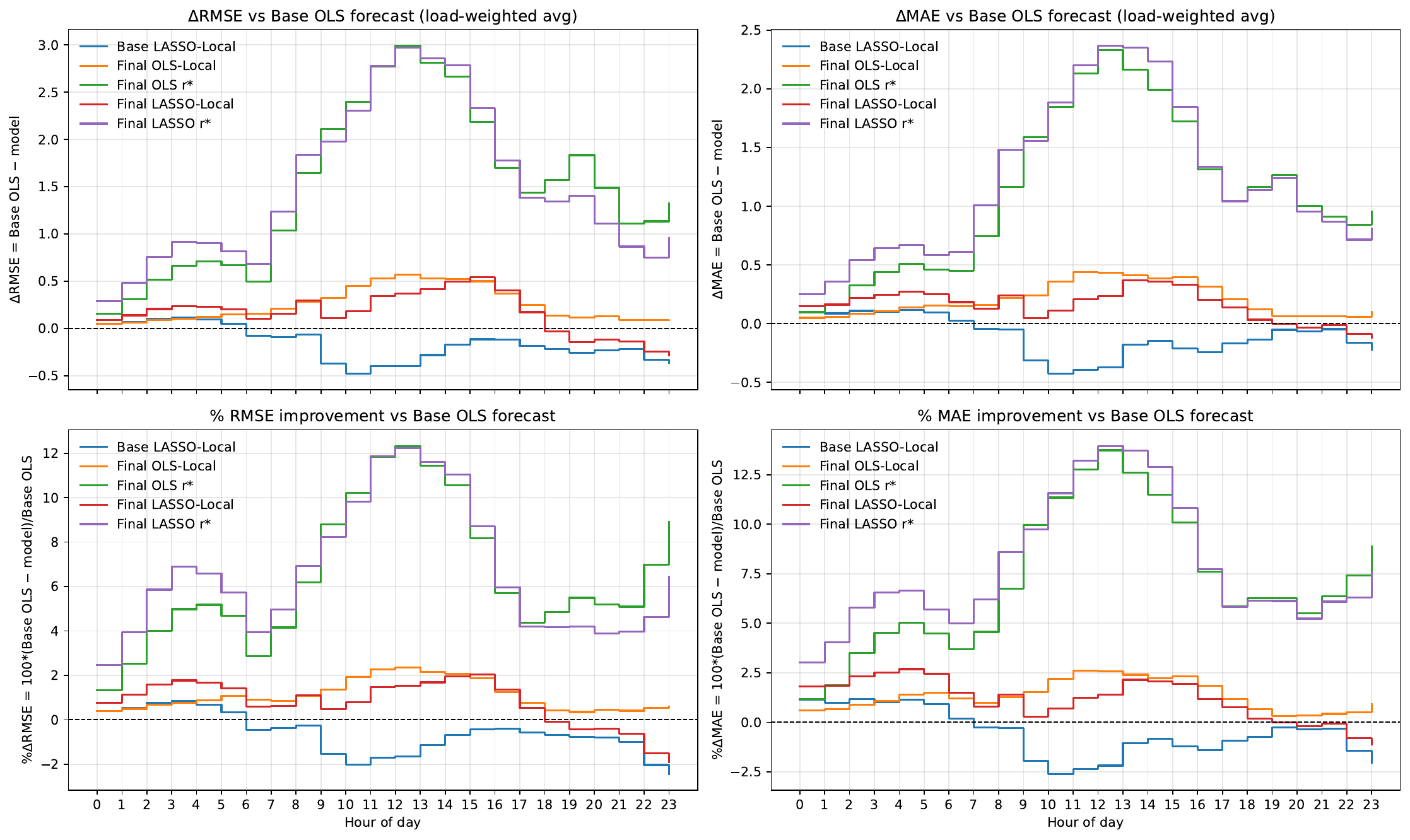}
  \caption{Hourly load-weighted RMSE and MAEs over the test sample are presented. The top panels show the differences in RMSE and MAE between selected models and the OLS Local benchmark (r = 0). The bottom panels report the corresponding percentage improvements relative to the benchmark, computed as $100\times(\text{base}-\text{model})/\text{base}$. Positive values indicate improved forecasting performance relative to the benchmark.}
  \label{fig:hcomp}
\end{figure}
We have implemented the NSTM models using OLS and Lasso and found that the results are the similar regardless of the estimation method. The results are quite surprising as the current EPF literature is overwhelmingly focused in BZN specific particualrs and frequently ignore the network effects. Across Europe, the NSTMs improve load-weighted RMSE by 7.2 percent and 7.3 percent for the Lasso and OLS methods, respectively. For MAE, the performance increase is even higher where the load-weighted outperformance of 9.4 percent and 8.2 percent, respectively, for Lasso and OLS. We also observe that the relationship between regional coverage ($\Delta r$) and incremental performance is not linear; instead, there seems to be an optimal neighborhod depth ($r^*$) at which performance seem to be at maximum unique for each BZN. The optimal neighborhood ball distances frequently fall in the vicinity between 2 and 4.  \vspace{.2cm}

If we look into the average hourly performances of the selected models including the best spatial model from each class relative to OLS based local model in the figure \ref{fig:hcomp}, The NSTM models with optimal distance ($r^*$ for the best weighted average performance) significantly outperform the local models across all hours of the day, highlighting the effectiveness of incorporating spatial information in electricity price forecasting. The findings are comprehensive across both estimation methods (OLS and Lasso), across BZNs and hours. The performance grain is particularly high during the peak hours of the day, when load levels and renewable generation variablility are highest. In figure \ref{fig:hcomp}, the spatial models perform with wide margins during the middle of the day, indicating the influence of renewable outputs and cross-border flows. During the early hours, the performance difference between the local and spatial models are relatively low compared to peak hours but highly significant nonetheless. Due to their efficiency in handling high-dimensional data, Lasso models perform better than their OLS counterparts during peak hours; on the other hand, OLS-based models seem to perform better at the end of the day. \vspace{.2cm}

\begin{figure}[H]
  \centering
  \captionsetup{justification=justified,singlelinecheck=false, font=footnotesize}
  \includegraphics[width=1\textwidth]{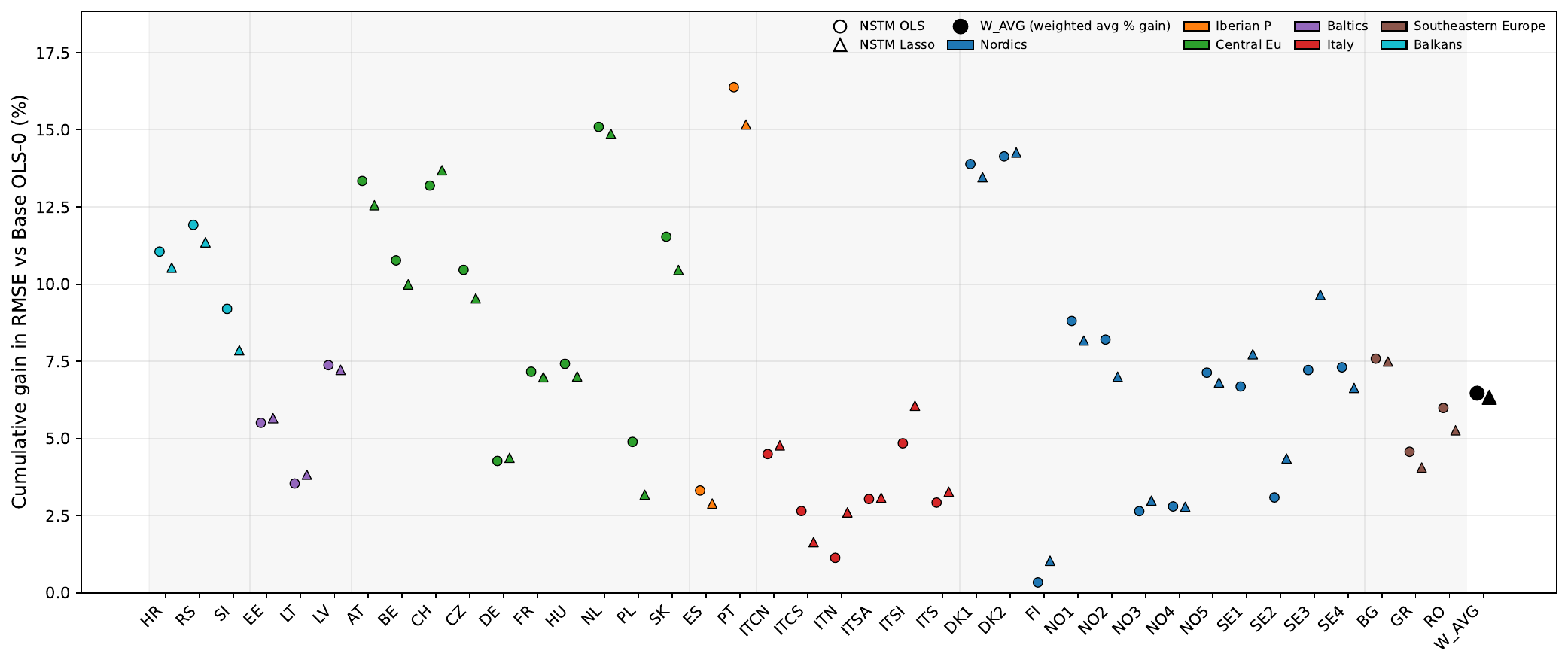}
  \caption{The figure shows the BZN-wise RMSE performance comparison of the best NSTM model from each class (OLS and Lasso) against the local model ($r=0$) across 39 European BZNs.}
  \label{fig:hcomp_bzn_rank_rmse}
\end{figure}

BZNs shows remarkable spatial correlations (Figure \ref{fig:winter-zoneblocks}), indicating that there are significant common drivers of price across the zones at least across clustered regions. The regional clustering is also prominent in the Day-Ahead prices but tends to change over time/seasons. This is also in line with the findings that the RMSE and MAE performance of the NSTM post-processed model varies across BZNs and by geographic clustering. We observe that the spatial model outperforms the pure local models across model classes, estimation regions, and clusterings.  In the figure \ref{fig:hcomp_bzn_rank_rmse}, the RMSE performances of various model classes and estimation methods, zone-wise and clustered, are presented. 

 The nature of the network and fundamental characteristics of the BZNs also play roles. For instance, the neighboring BZNs in the Iberian Peninsula, such as Portugal (PT) and Spain (ES), are physically very close, but in terms of spatial model performance, Portugal's NSTM model performs much better than Spain's. The Spanish bidding zone has load and generation profiles that are disproportionately higher than Portugal's, so Spain, in general, exerts more meaningful influence on Portugal. Also, smaller BZNs with rich connectivity genrally can take advantage of the NSTM. For instances, the Danish bidding zones show noteworthy performance in the NSTM as both Danish bidding zones are closely interconnected with mainland Europe and other Nordic countries. The performance of larger bidding zones, such as France, Poland, and Germany, lags behind that of medium and smaller sized bidding zones in central Europe. The relatively lower performances of some of the peripheral Nordics, Italians, and Baltic countries are remarkable. It is worth noting that incremental performance generally diminishes as the neighborhood radius increases. It supports our hypothesis that neither pure local nor full-on global is the optimal design choice when incorporating spatial models in electricity price forecasting. It also supports our conjecture that there exists an optimal distance $r^{*}$ between the pure local ($r=0$) and maximum coverage \textbf{max($r$)}, where the optimal model will dominate the forecasting errors over models with non-optimal neighborhood coverage. 
\begin{figure}[H]
  \centering
  \captionsetup{justification=justified,singlelinecheck=false, font=footnotesize}
  \includegraphics[width=.8\textwidth]{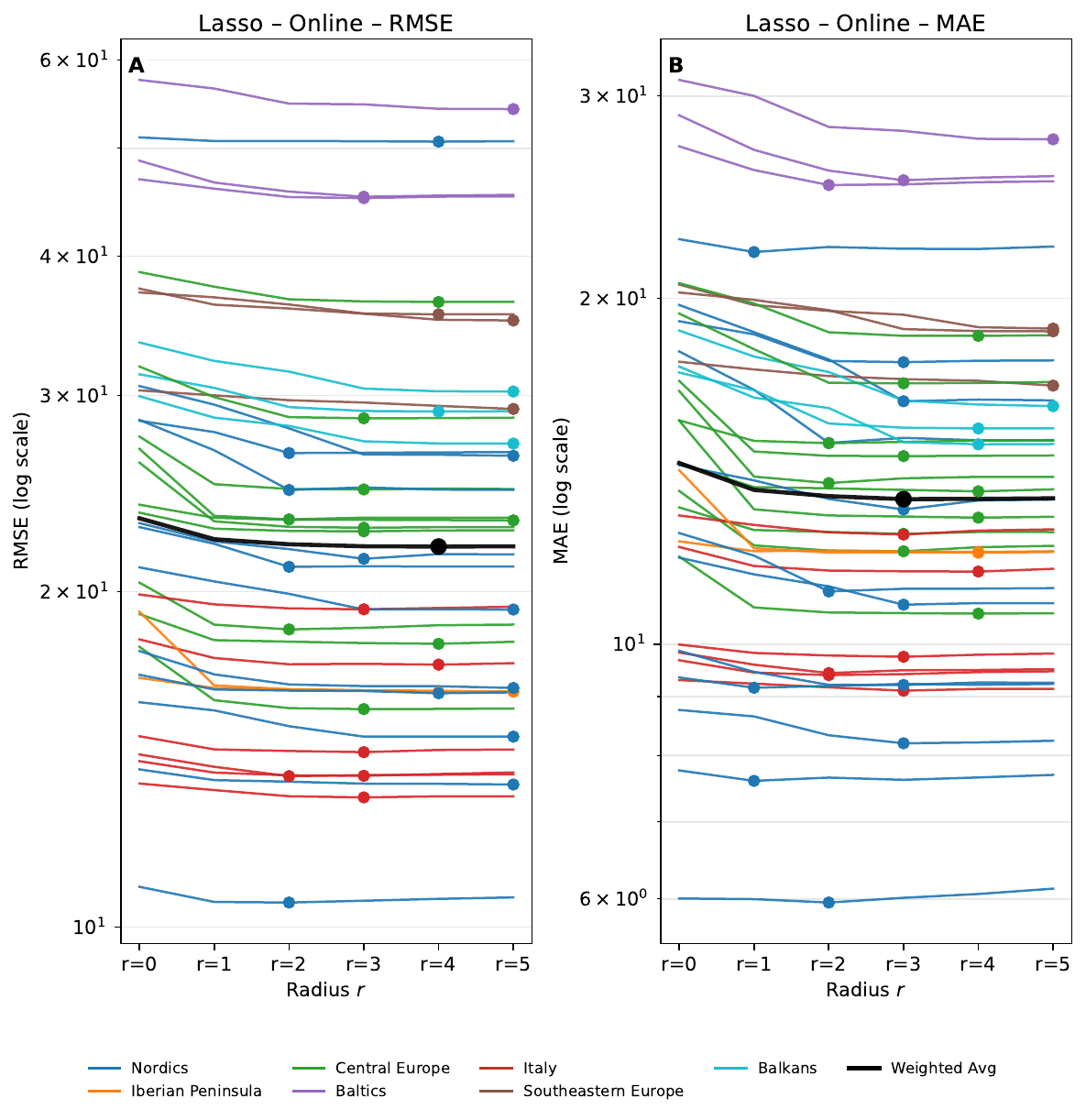}
  \caption{The figure shows RMSE and MAE of NSTM model classes from the test sample. The RMSE and MAE with optimal distance ($r^*$) of the NSTM are shown by individual lines for all bidding zones (BZNs). The BZNs are colour coded to differentiate across different geographic clusters.}
  \label{fig:perf_optimumr}
\end{figure}
The bidding zones interconnect uniquely with their neighbors, and some zones at the edge have only one-sided neighbors. In this current setup, it takes 6 to 11 steps for a bidding zone to establish a full Europe-wide connection. Although there are idiosyncrasies across bidding zones or even regional clusters, we observe that, in most bidding zones, the spatial models achieve optimality within 2 to 4 distance ($r^*$). Although the optimal regional coverage with respect to RMSE and MAE varies slightly, it is close in most cases. Individual BZN-wise model performances (both RMSE and MAE) also support this observation. The figure \ref{fig:perf_optimumr} also shows that most bidding zones achieve most of their performance using information from closer neighbors. As we increase the number of neighbors by expanding the neighborhood ball with a larger $r$, the improvement in performance tapers off. In the post processed final model, the majority of the spatial model's outperformance stems from direct spatial spillovers from neighbors. 

\begin{figure}[H]
  \centering
  \captionsetup{justification=justified,singlelinecheck=false, font=footnotesize}
  \includegraphics[width=.9\textwidth]{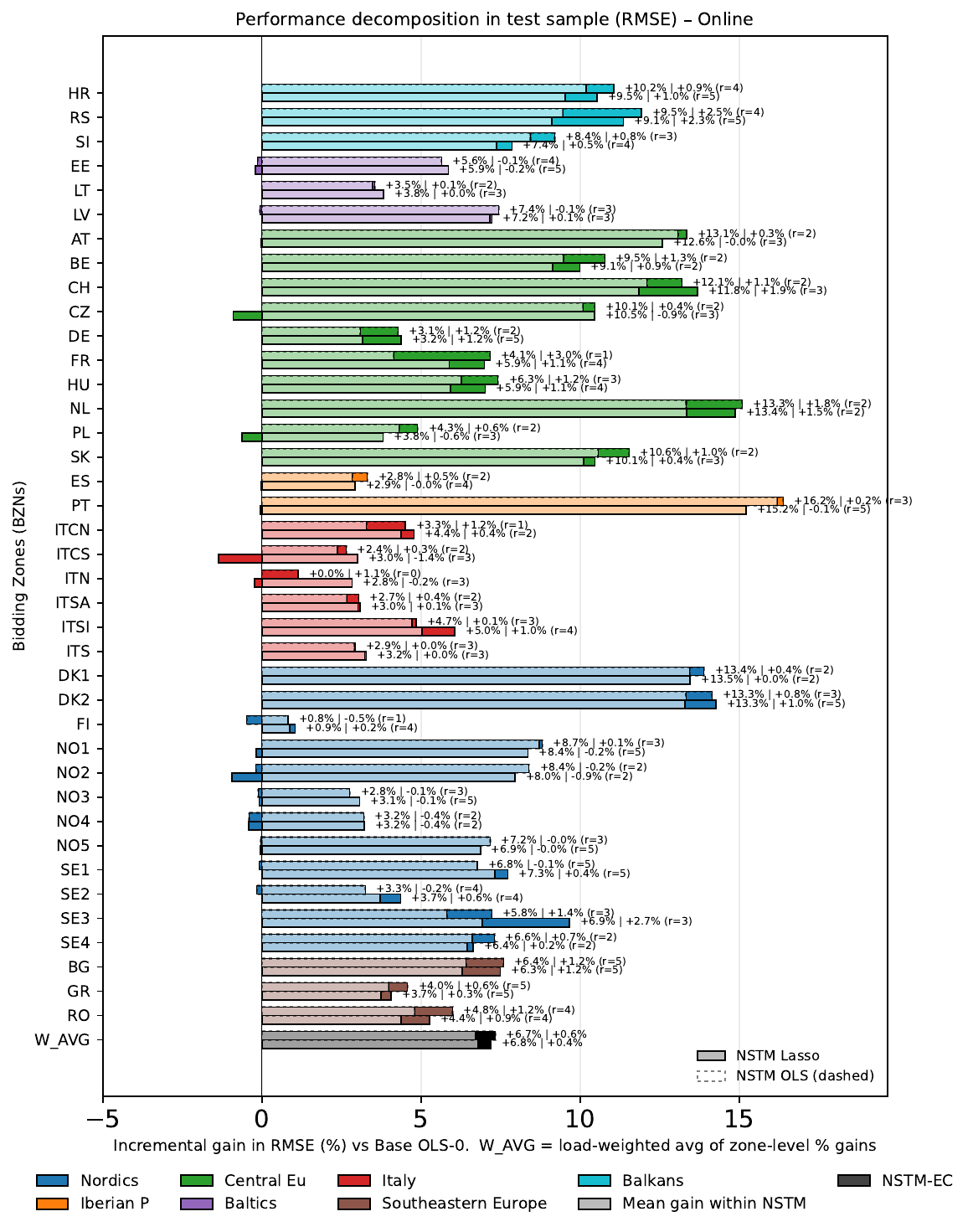}
  \caption{The performance gains relative to the baseline are decomposed into contributions from the mean component (NSTM) and the error-correction term. Improvements are computed using the zone-specific optimal distance $r^\ast$. The aggregate (W\_AVG) reports the load-weighted average of zone-level percentage gains, $\sum_k w_k g_k / \sum_k w_k$.}
  \label{fig:perf_perfattribution_RMSE_vs_r0}
\end{figure} 
The figure \ref{fig:perf_perfattribution_RMSE_vs_r0} provides the performance attribution of the mean model plus error correction extension for RMSE. For the RMSE metric of Lasso-based models, the total 7.2\%  weighted average outperforms the 6.8 \%  from the Mean NSTM component and the 0.4\%  from the error correction step. For OLS, the decomposition is 6.7\%  and 0.6\% , respectively. For the MAE of Lasso-based models, 8.3\% and 1.1\% of the outperformance comes from the mean and residual adjustment components, respectively. On the other hand, for OLS, the same trend was observed, where the majority of the outperformance is owing to the mean model component. On an individual BZN basis, NSTM outperforms the pure local model across the board. The Diebold-Mariano (DM) test results for each individual bidding zone in the test sample  also confirm the spatial models' outperformance. \vspace{.2cm}

\begin{figure}[H]
  \centering
  \captionsetup{justification=justified,singlelinecheck=false, font= footnotesize}
  \includegraphics[width=1\textwidth]{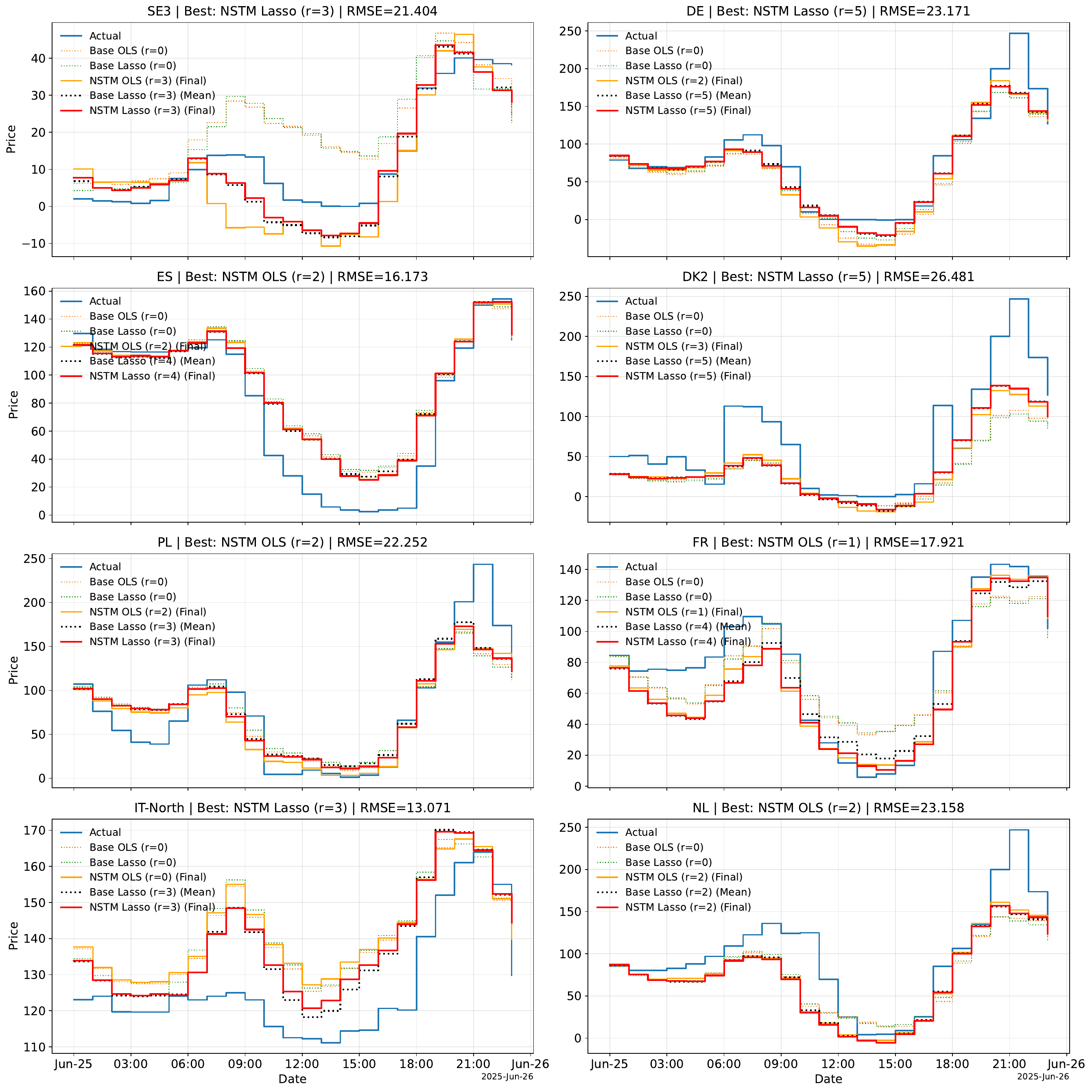}
  \caption{The figure compares actual prices with forecasts from Local OLS, Local Lasso, and their respective spatial NSTMs, evaluated at the optimal spatial distance ($r^*$) for selected bidding zones. Each panel identifies the top-performing model and its RMSE for each zone.}
  \label{fit_nstm}
\end{figure}

In this networked, spatial-temporal backdrop, the separate hourly models' residuals are centered arouned mean zero. Despite that, the residuals do correlate highly across hours and regional clusters. Even after implementing the base Networked Spatio-Temporal Model (NSTM), the same-day residuals across hourly models show considerable correlation, especially between neighboring hours. It indicates that the estimated mean model with a linear-only specification may be well calibrated across days, but it is subject to the latent factors across intra-day hours. On top of that, the clustered spatial structure also shows a seasonal patterns. To illustrate the forecasting fit of the models, we present a stylized example of 8 selected bidding zones for a 24-hour model on June 25, 2025 in Figure \ref{fit_nstm}. The figure compares the forecasting fit of selected NSTM and Local Models. The NSTM models with optimal distance ($r^*$) significantly outperform the local models across all hours of the day, revealing the effectiveness of incorporating spatial information in electricity price forecasting. The bidding zones have significant idiosyncrasies, so there is no single global model suitable for all bidding zones; each must be customized to account for the myriad effects that dominate the price process in that zone. As we already suspected, the nature of the indtra-day dynamics is somewhat complex and plausibly non-linear; approximating with a linear model class improves the result but there are space left for further improvements.

\begin{figure}[H]
    \centering
    \captionsetup{justification=justified,singlelinecheck=false, font=footnotesize}
    \includegraphics[width=1.0\textwidth]{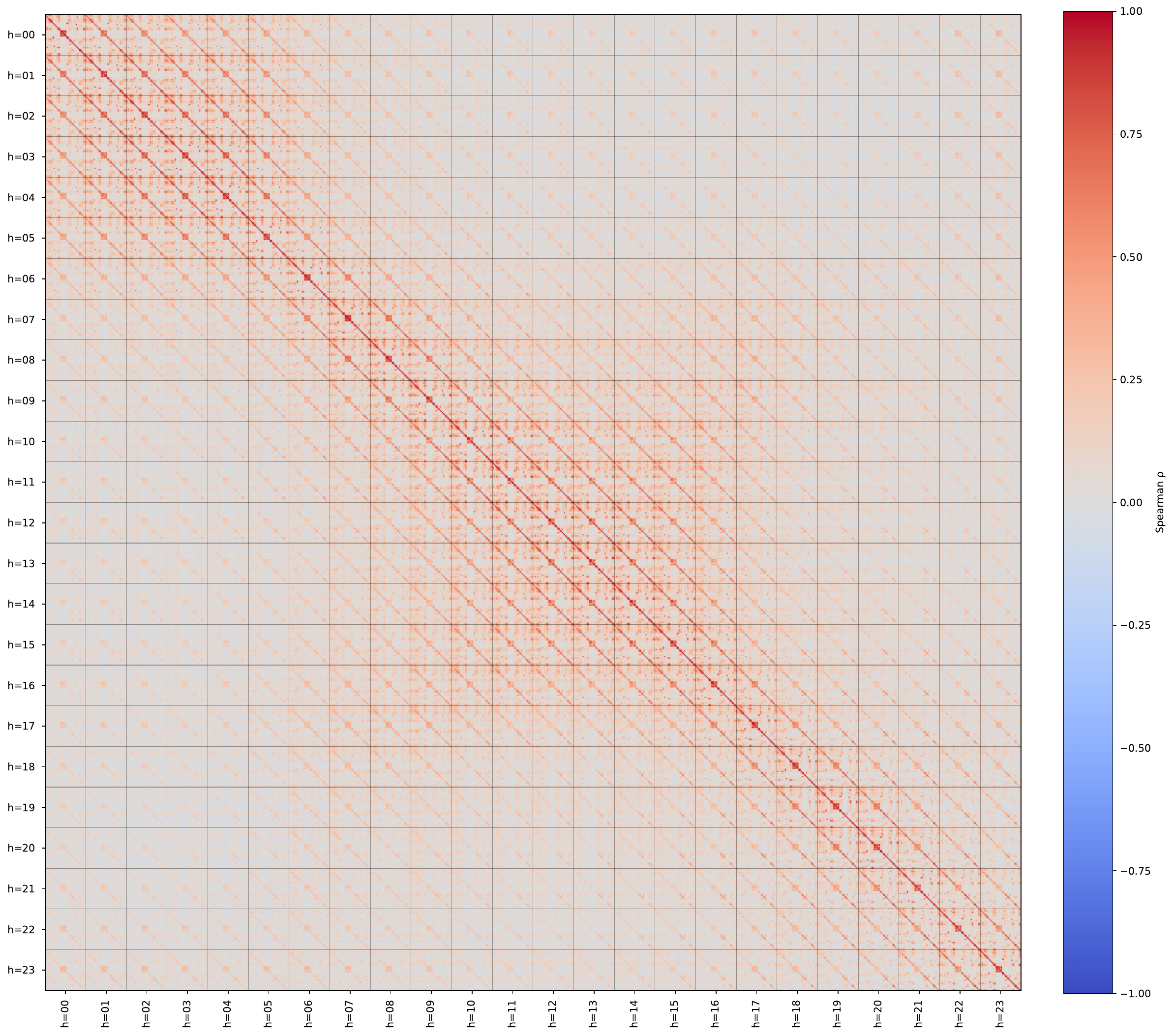}
    \caption{ Spearman rank correlation of hourly model residuals during the test period (Oct~2023--Sep~2025) across hours and bidding zones. 
    }
\label{fig:hourzone-residuals}
\end{figure}

Figure \ref{fig:hourzone-residuals} shows the spatio-temporal residual strutures where the temporal dimension corresponds to intraday hours, and the spatial dimension corresponds to the bidding zones. The figure is constructed from hourly residuals of the NSTM–LASSO-3 final model (with post-processing error correcion) across all bidding zones and hours of the day, resulting in a $(24K \times 24K)$ correlation matrix. It is organized into hour blocks ($24\times24$), where each block represents zone-by-zone correlations across the $K$ bidding zones. Low off-diagonal correlations indicate that most spatio-temporal dependence in electricity prices is captured by the forecasting model. Here, it still maintains significant correlation specially in the diagonals direction although the post-processing do some improvements. Generally for NSTMS, the intraday correlations in the off-diagonal blocks are considerably reduced for the concurrent residuals across the board, at lagged basis, almost none. This demonstration is also a hallmark of the existence of a spatiotemporal process, in which the intraday structure still contains information about a relationship that is much more complex.

\section{Conclusion and outlook} European electricity price forecasting models have evolved as the grid has become increasingly interconnected and renewables have come to dominate the generation profiles, making forecasting more complex. While well-defined linear mechanisms explain general electricity price expectations, short-term divergences and regime changes do still occur, making forecasting challenging. These divergences may result from non-linear spillovers from neighbors, higher-order effects, complex interactions, uncertainty, or supply-demand shocks. Our proposed NSTM provides a flexible, high-resolution modeling approach that integrates spatial information and can track explainable performance differentials as models are nested. The spatial impact is comprehensive and systemic. Given that we have a consistent estimator in a linear modeling approach, we can estimate the sources of direct linear impact on model performance. On the other hand, while we assume the first-stage residuals are reasonably well behaved for hourly models, they exhibit significant cross-hour and cross-BZN dynamics. Although we tried to capture some of the short-term divergences through the post-processing stage, it does not capture the full story. The electricity market mechanism ensures that prices have finite moments: they are capped and regulated, and bounded by behavioral constraints, while demand is finite. So the mean level is generally dominated by seasonalities and fundamental factors, while short- to medium-term divergences are complex and exhibit high nonlinearity, which we can address with more complex models that effectively capture intraday dynamics and the general trajectory of the mean level. That is why the spatio-temporal elements are so critical in modeling all stages of forecasting (short- and long-term).  We have kept the model specification similar across hours and BZNs to compare spatial effects from a uniform-modeling perspective. It is worth noting that, although we have used a high-resolution hourly model structure, the models can be defined separately for each hour and each BZN, providing a high level of flexibility in terms of model specification and classes. In this backdrop, incorporating nonlinear components via more advanced models or machine learning/deep learning algorithms opens a new horizon for modeling a large number of price objects in an integrated manner. The network-graph structure is thus a fundamental reality of the European electricity market, as demonstrated in this paper. The strong results of the NSTM models show the need for a more comprehensive feature space. Most of the NSTM model's performance stems from direct spillover effects. To a lesser extent, it comes from the post-processing error-correction model. We admit there is room to specify latent factors for proper shock modeling or to account for pervasive non-linearity across BZNs. This will definitely yield more precise, sharper forecasts. Further, the NSTM can also be extended to model nonlinearity and serve as a backdrop for more complex model construction. From a linear modeling perspective, the comprehensive linear spillover effect from the neighboring bidding zones is a testament to the fact that, without the interconnected networked grid information, any model will suffer from omitted-variable bias.  \vspace{.2cm}

 \vspace{2 cm}

%
\subsection*{\textbf{Declaration of competing interest:}}
The authors declare that they have no known competing financial interests or personal relationships that could have influenced the work reported in this paper. 

\subsection*{\textbf{Acknowledgments:}}

This research was partially funded in the course of TRR 391 Spatio-temporal Statistics for the Transition of Energy and Transport (520388526) by the Deutsche Forschungsgemeinschaft (DFG, German Research Foundation).

\newpage
\section{Appendix}

\begin{figure}[H]
  \centering
  \captionsetup{justification=justified,singlelinecheck=false, font=footnotesize}
  \includegraphics[width=.9\textwidth]{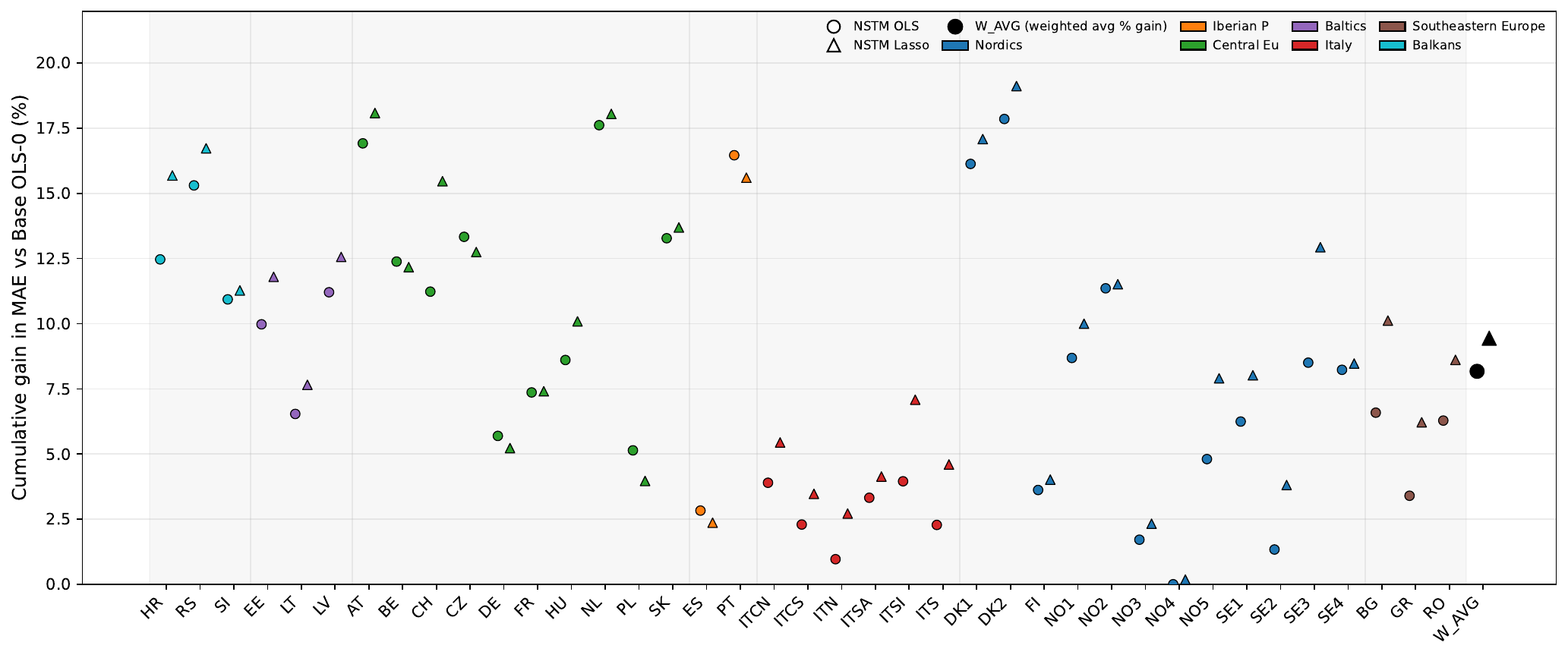}
  \caption{The figure shows the BZN-wise MAE performance comparison of the best NSTM model from each class (OLS and Lasso) against the local model ($r=0$) across 39 European BZNs.}
  \label{fig:hcomp_bzn_rank_mae}
\end{figure}

\begin{figure}[H]
  \centering
  \captionsetup{justification=justified,singlelinecheck=false, font=footnotesize}
  \includegraphics[width=.7\textwidth]{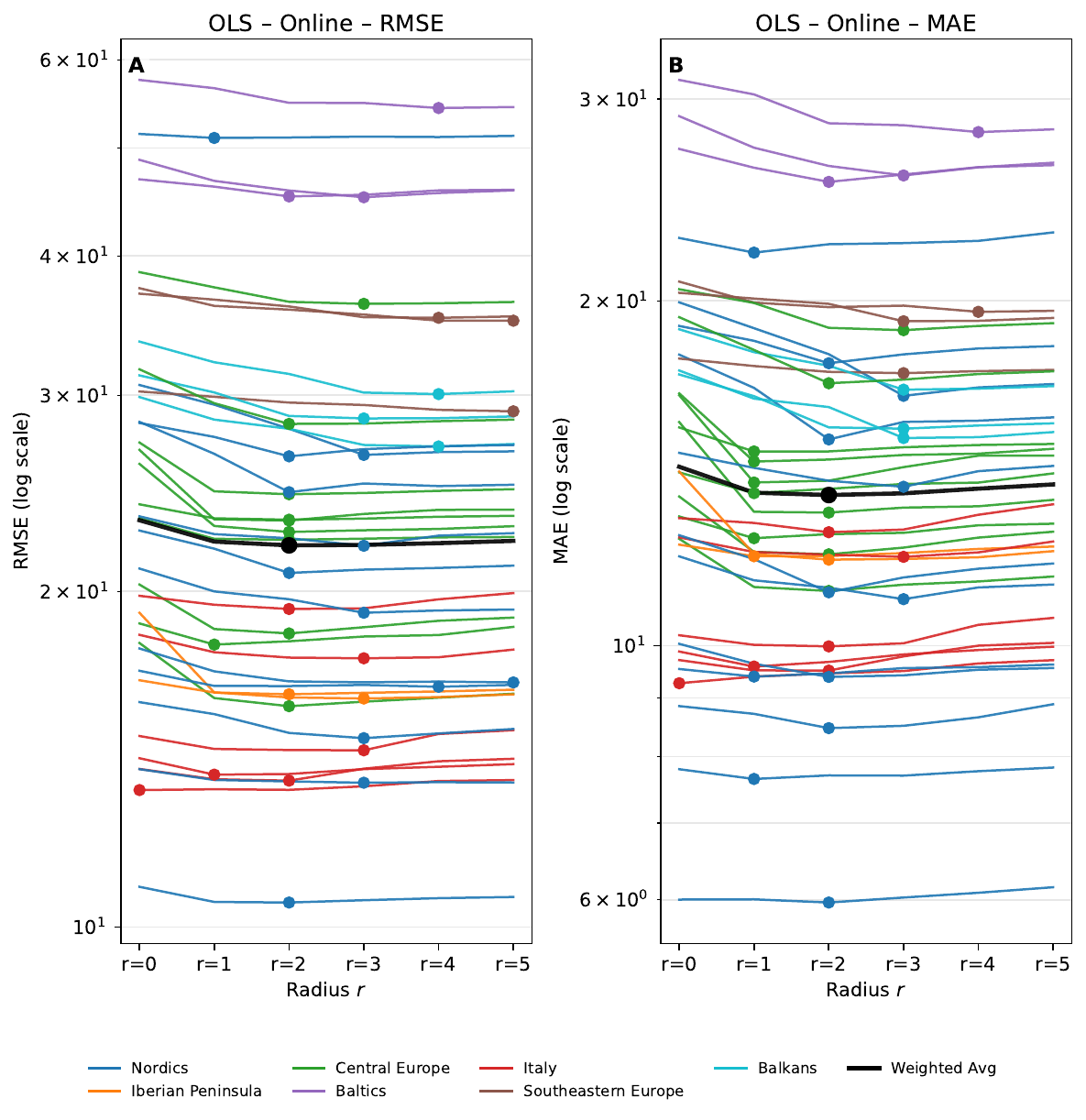}
  \caption{The figure shows RMSE and MAE of NSTM model classes from the test sample spanning Oct 2023--Sep 2025. The RMSE and MAE with optimal distance ($r^*$) of the NSTM are shown by individual lines for all bidding zones (BZNs).}
  \label{fig:perf_optimumr_ols}
\end{figure}

\begin{figure}[H]
  \centering
  \title{\textbf{MAE - NSTM}, test sample}
  \captionsetup{justification=justified,singlelinecheck=false, font= footnotesize}
  \includegraphics[width=1\textwidth]{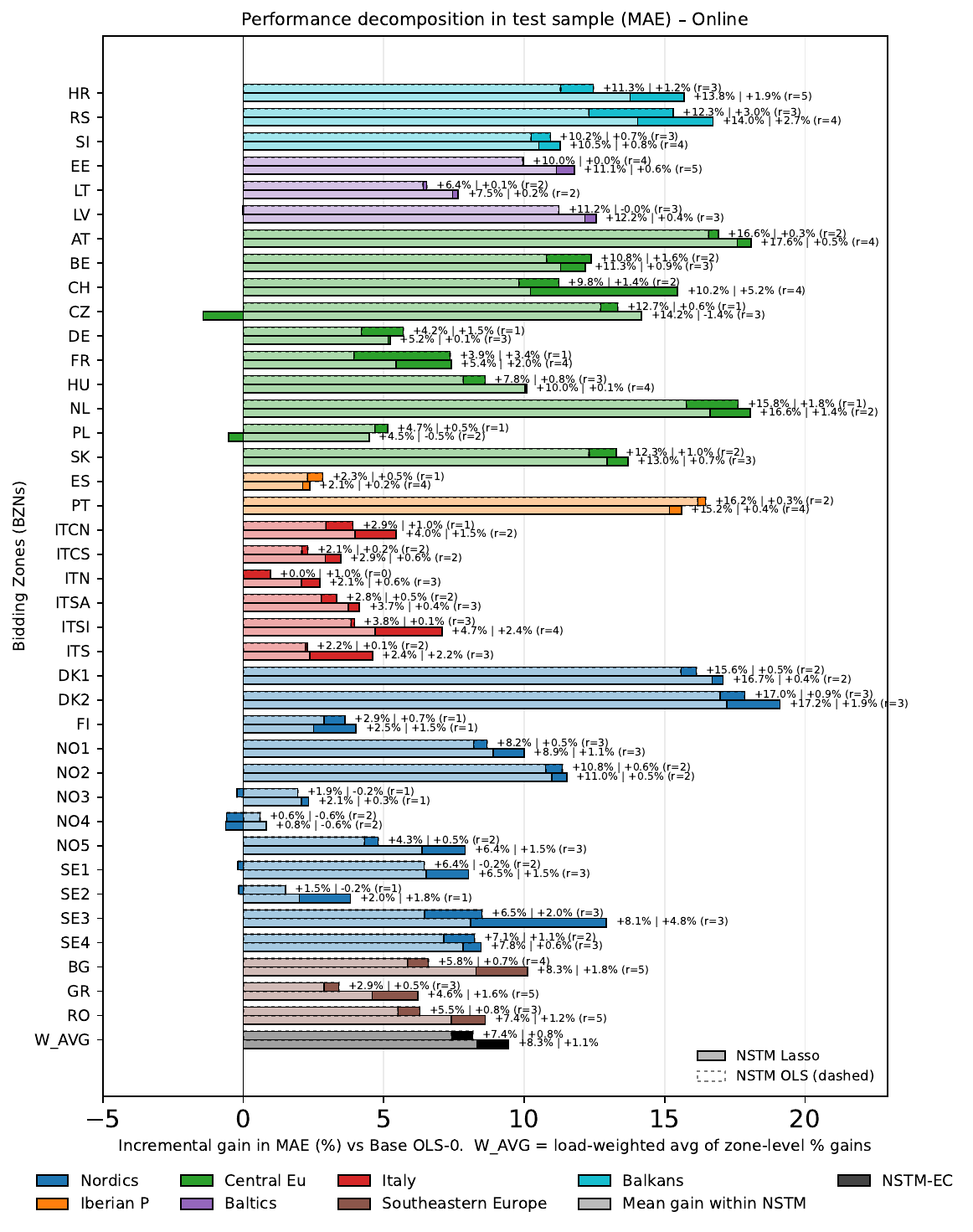}
  \caption{MAE performance decomposition of the NSTM model over the test period (Oct~2023--Sep~2025). Total performance gains relative to the baseline are decomposed into contributions from the mean component (NSTM) and the error-correction term. Improvements are computed using the zone-specific optimal distance $r^\ast$. The aggregate (W\_AVG) reports the load-weighted average of zone-level percentage gains, $\sum_k w_k g_k / \sum_k w_k$.}
  \label{perf_perfattribution_mae_vs_r0}
\end{figure}

\begin{figure}[H]
  \centering
  \captionsetup{justification=justified,singlelinecheck=false, font=footnotesize}
  \includegraphics[width=1\textwidth]{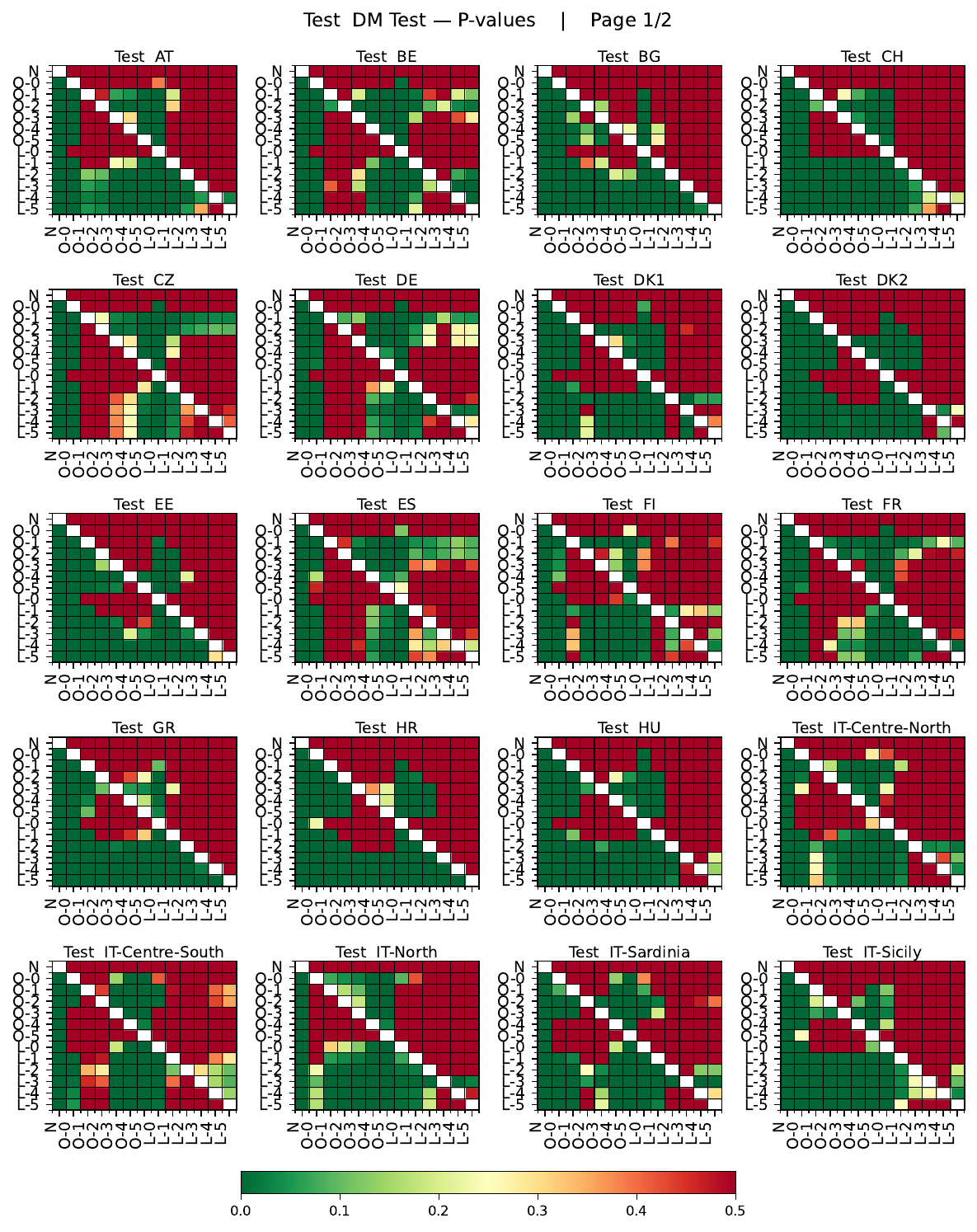}
  \caption{The figure shows the Diebold-Mariano (DM Test) results corresponding to the test sample spanning Oct 2023--Sep 2025. The figure includes the first 20 bidding zones' test results for all final NSTM model classes.}
  \label{fig:perfDM1}
\end{figure}

\begin{figure}[H]
  \centering
  \captionsetup{justification=justified,singlelinecheck=false, font=footnotesize}
  \includegraphics[width=1\textwidth]{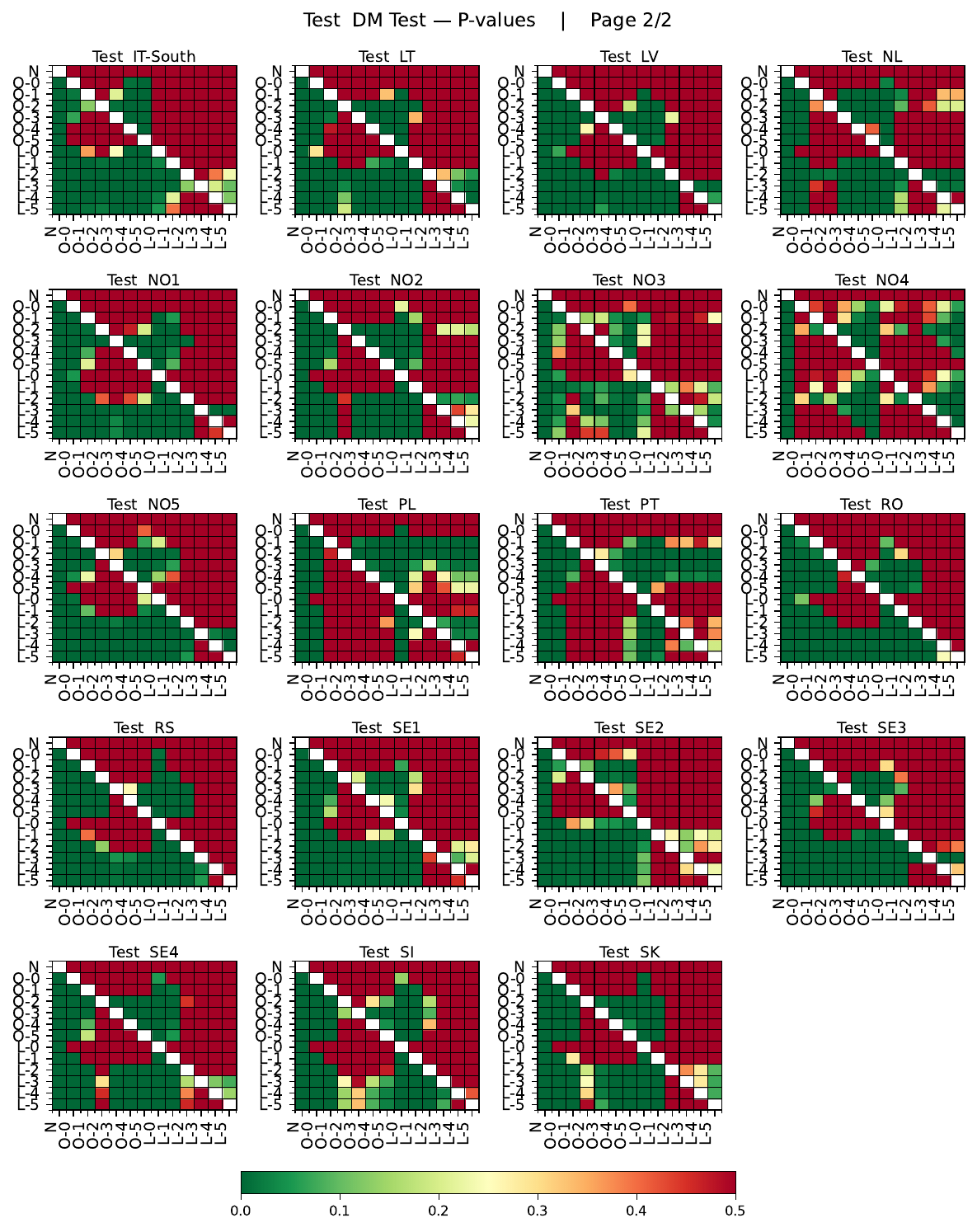}
  \caption{The figure shows the Diebold-Mariano (DM Test) results corresponding to the test sample spanning Oct 2023--Sep 2025. The figure includes the remaining 19 bidding zones' test results for all final NSTM model classes.}
  \label{fig:perfDM2}
\end{figure}

\begin{table} [H]
\centering
\scriptsize
\title{Summary of model performances across BZNs, Model classes, and estimation methods}
\setlength{\tabcolsep}{3pt}
\renewcommand{\arraystretch}{0.8}
\begin{tabular}{|l|rrrr|rr|rrrr|!{\vrule width 0.5pt}rrrr|rr|}
\toprule
Family & \multicolumn{4}{c|}{Optimal $r$} & \multicolumn{2}{c|}{OLS (Base)} & \multicolumn{4}{c|}{NSTM-OLS} & \multicolumn{4}{c|}{NSTM-LASSO} & \multicolumn{2}{c|}{Naive} \\
\cmidrule(lr){1-17}
Metric & \multicolumn{2}{c}{RMSE} & \multicolumn{2}{c|}{MAE} & \multicolumn{1}{c}{RMSE} & \multicolumn{1}{c|}{MAE} & \multicolumn{2}{c}{RMSE} & \multicolumn{2}{c|}{MAE} & \multicolumn{2}{c}{RMSE} & \multicolumn{2}{c|}{MAE} & \multicolumn{1}{c}{RMSE} & \multicolumn{1}{c|}{MAE} \\
\cmidrule(lr){1-17}
BZN/Col. & \rotatebox{0}{O} & \rotatebox{0}{L} & \rotatebox{0}{O} & \rotatebox{0}{L} & \rotatebox{0}{O-0} & \rotatebox{0}{O-0} & \rotatebox{0}{O-0} & \rotatebox{0}{O ($r^\ast$)} & \rotatebox{0}{O-0} & \rotatebox{0}{O ($r^\ast$)} & \rotatebox{0}{L-0} & \rotatebox{0}{L ($r^\ast$)} & \rotatebox{0}{L-0} & \rotatebox{0}{L ($r^\ast$)} & \rotatebox{0}{Naive} & \rotatebox{0}{Naive} \\ \cmidrule(lr){1-17}
AT & 2 & 3 & 2 & 4 & 26.10 & 15.73 & 26.03 & 22.62 & 15.67 & 13.07 & 26.11 & 22.82 & 15.65 & 12.88 & 35.25 & 21.58 \\
BE & 2 & 2 & 2 & 3 & 20.55 & 13.71 & 20.29 & 18.34 & 13.50 & 12.01 & 20.37 & 18.50 & 13.59 & 12.04 & 36.08 & 24.42 \\
BG & 5 & 5 & 4 & 5 & 37.87 & 20.94 & 37.42 & 34.99 & 20.79 & 19.56 & 37.41 & 35.03 & 20.56 & 18.82 & 45.88 & 25.52 \\
CH & 2 & 3 & 2 & 4 & 18.18 & 12.57 & 17.98 & 15.78 & 12.40 & 11.16 & 17.84 & 15.69 & 11.92 & 10.63 & 24.27 & 15.42 \\
CZ & 2 & 3 & 1 & 3 & 27.31 & 16.71 & 27.20 & 24.45 & 16.60 & 14.48 & 27.55 & 24.70 & 16.95 & 14.58 & 36.94 & 23.39 \\
DE & 2 & 5 & 1 & 3 & 24.23 & 13.16 & 23.95 & 23.20 & 12.96 & 12.41 & 23.94 & 23.17 & 13.15 & 12.47 & 42.64 & 27.37 \\
DK1 & 2 & 2 & 2 & 2 & 28.52 & 18.05 & 28.39 & 24.55 & 17.95 & 15.14 & 28.51 & 24.68 & 17.98 & 14.97 & 45.89 & 30.37 \\
DK2 & 3 & 5 & 3 & 3 & 30.89 & 20.11 & 30.64 & 26.52 & 19.94 & 16.52 & 30.59 & 26.48 & 19.73 & 16.27 & 47.25 & 31.64 \\
EE & 4 & 5 & 4 & 5 & 57.47 & 31.18 & 57.54 & 54.30 & 31.18 & 28.07 & 57.58 & 54.21 & 30.98 & 27.51 & 68.49 & 40.98 \\
ES & 2 & 4 & 1 & 4 & 16.73 & 12.32 & 16.65 & 16.17 & 12.25 & 11.97 & 16.73 & 16.24 & 12.29 & 12.03 & 33.41 & 22.40 \\
FI & 1 & 4 & 1 & 1 & 51.22 & 22.86 & 51.47 & 51.05 & 22.69 & 22.03 & 51.14 & 50.69 & 22.52 & 21.94 & 66.92 & 36.42 \\
FR & 1 & 4 & 1 & 4 & 19.30 & 14.67 & 18.72 & 17.92 & 14.16 & 13.59 & 19.09 & 17.96 & 14.38 & 13.58 & 33.23 & 23.38 \\
GR & 5 & 5 & 3 & 5 & 30.41 & 17.90 & 30.23 & 29.02 & 17.81 & 17.29 & 30.31 & 29.17 & 17.61 & 16.79 & 39.47 & 23.09 \\
HR & 4 & 5 & 3 & 5 & 33.81 & 19.11 & 33.51 & 30.07 & 18.89 & 16.73 & 33.47 & 30.25 & 18.75 & 16.11 & 40.50 & 23.56 \\
HU & 3 & 4 & 3 & 4 & 39.14 & 20.63 & 38.69 & 36.23 & 20.47 & 18.85 & 38.71 & 36.39 & 20.62 & 18.55 & 46.06 & 25.95 \\
IT-CN & 1 & 2 & 1 & 2 & 14.34 & 9.98 & 14.17 & 13.70 & 9.88 & 9.59 & 14.29 & 13.66 & 9.83 & 9.43 & 19.61 & 12.82 \\
IT-CS & 2 & 3 & 2 & 2 & 13.90 & 9.73 & 13.86 & 13.53 & 9.71 & 9.51 & 14.09 & 13.67 & 9.68 & 9.40 & 20.64 & 13.35 \\
IT-N & 0 & 3 & 0 & 3 & 13.42 & 9.36 & 13.27 & 13.27 & 9.27 & 9.27 & 13.45 & 13.07 & 9.30 & 9.11 & 19.21 & 12.61 \\
IT-Sar & 2 & 3 & 2 & 3 & 19.89 & 12.99 & 19.82 & 19.29 & 12.92 & 12.56 & 19.88 & 19.28 & 12.94 & 12.45 & 27.85 & 16.51 \\
IT-Sic & 3 & 4 & 3 & 4 & 18.31 & 12.45 & 18.28 & 17.42 & 12.43 & 11.95 & 18.12 & 17.20 & 12.15 & 11.56 & 24.54 & 15.44 \\
IT-S & 3 & 3 & 2 & 3 & 14.84 & 10.22 & 14.84 & 14.41 & 10.21 & 9.99 & 14.83 & 14.35 & 9.99 & 9.75 & 21.80 & 13.91 \\
LT & 2 & 3 & 2 & 2 & 46.90 & 27.18 & 46.86 & 45.24 & 27.14 & 25.40 & 46.89 & 45.10 & 27.12 & 25.10 & 62.80 & 39.91 \\
LV & 3 & 3 & 3 & 3 & 48.76 & 28.98 & 48.79 & 45.16 & 28.99 & 25.73 & 48.73 & 45.23 & 28.86 & 25.34 & 62.89 & 40.05 \\
NL & 2 & 2 & 1 & 2 & 27.28 & 16.85 & 26.80 & 23.16 & 16.54 & 13.88 & 26.86 & 23.22 & 16.61 & 13.81 & 39.93 & 25.89 \\
NO1 & 3 & 5 & 3 & 3 & 20.99 & 12.02 & 20.97 & 19.14 & 11.97 & 10.98 & 21.03 & 19.27 & 11.89 & 10.82 & 29.03 & 16.94 \\
NO2 & 2 & 2 & 2 & 2 & 22.64 & 12.56 & 22.68 & 20.78 & 12.48 & 11.13 & 22.85 & 21.05 & 12.49 & 11.11 & 31.42 & 17.66 \\
NO3 & 3 & 5 & 1 & 1 & 13.84 & 7.78 & 13.85 & 13.47 & 7.80 & 7.65 & 13.85 & 13.42 & 7.76 & 7.60 & 20.70 & 12.13 \\
NO4 & 2 & 2 & 2 & 2 & 10.82 & 5.97 & 10.86 & 10.52 & 6.00 & 5.97 & 10.87 & 10.52 & 6.00 & 5.96 & 16.43 & 8.21 \\
NO5 & 3 & 5 & 2 & 3 & 15.91 & 8.90 & 15.91 & 14.77 & 8.86 & 8.47 & 15.91 & 14.82 & 8.76 & 8.20 & 21.17 & 12.42 \\
PL & 2 & 3 & 1 & 2 & 23.40 & 15.58 & 23.26 & 22.25 & 15.50 & 14.78 & 23.54 & 22.65 & 15.66 & 14.96 & 36.82 & 24.47 \\
PT & 3 & 5 & 2 & 4 & 19.17 & 14.23 & 19.14 & 16.03 & 14.19 & 11.89 & 19.18 & 16.26 & 14.17 & 12.01 & 33.06 & 22.27 \\
RO & 4 & 4 & 3 & 5 & 37.45 & 20.48 & 37.00 & 35.20 & 20.32 & 19.19 & 37.11 & 35.47 & 20.23 & 18.72 & 46.50 & 25.87 \\
RS & 4 & 5 & 3 & 4 & 30.64 & 17.92 & 29.89 & 26.99 & 17.38 & 15.18 & 29.95 & 27.16 & 17.44 & 14.92 & 36.75 & 21.50 \\
SE1 & 5 & 5 & 2 & 3 & 17.76 & 10.02 & 17.78 & 16.57 & 10.04 & 9.39 & 17.69 & 16.39 & 9.86 & 9.21 & 26.08 & 14.76 \\
SE2 & 4 & 4 & 1 & 1 & 16.95 & 9.53 & 16.98 & 16.43 & 9.54 & 9.40 & 16.84 & 16.21 & 9.35 & 9.16 & 26.60 & 15.15 \\
SE3 & 3 & 3 & 3 & 3 & 23.69 & 15.04 & 23.36 & 21.98 & 14.73 & 13.76 & 23.04 & 21.40 & 14.31 & 13.10 & 36.71 & 22.81 \\
SE4 & 2 & 2 & 2 & 3 & 28.53 & 19.22 & 28.33 & 26.44 & 19.01 & 17.64 & 28.47 & 26.63 & 19.10 & 17.60 & 45.20 & 29.71 \\
SI & 3 & 4 & 3 & 4 & 31.49 & 17.37 & 31.25 & 28.59 & 17.25 & 15.47 & 31.34 & 29.02 & 17.23 & 15.41 & 40.17 & 23.36 \\
SK & 2 & 3 & 2 & 3 & 31.96 & 19.54 & 31.65 & 28.27 & 19.35 & 16.95 & 31.85 & 28.62 & 19.40 & 16.87 & 39.12 & 24.63 \\
\midrule
W. Avg. & - & - & - & - & 23.42 & 14.53 & 23.17 & 21.88 & 14.33 & 13.42 & 23.27 & 21.92 & 14.37 & 13.33 & 36.23 & 23.15 \\
\bottomrule
\end{tabular}
\captionsetup{justification=justified,singlelinecheck=false,font=footnotesize}
\caption{The table presents bidding-zone level forecast accuracy over the test sample. The table reports RMSE and MAE for baseline and spatial forecasting models. OLS (Base) corresponds to the local specification ($r=0$). NSTM--OLS and NSTM--LASSO denote spatial models estimated using OLS and LASSO, respectively. Columns ``O'' and ``L'' indicate OLS and Lasso based models and the $r^\ast$ denotes optimal spatial radius. The final row reports load-weighted averages across bidding zones.}
\label{tab:bzn_rmse_mae_compact_with_base_ols}
\end{table}

\begin{figure}[H]
    \centering
    \captionsetup{justification=justified,singlelinecheck=false, font=footnotesize}
    \includegraphics[width=1.0\textwidth]{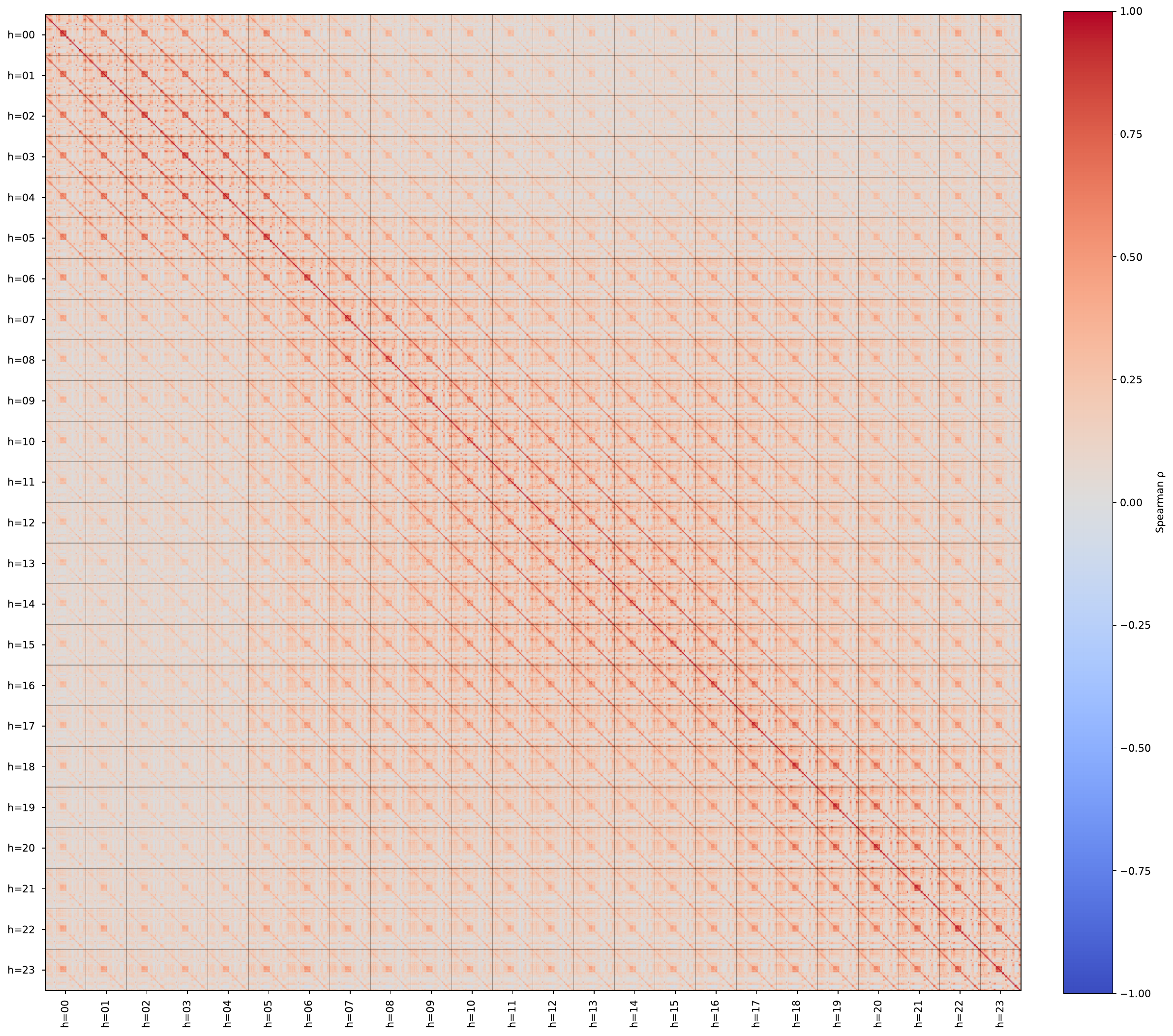}
    \caption{ Spearman rank correlation of hourly model residuals during the train period (Oct~2020--Sep~2023) across hours and bidding zones. As shown in figure, the temporal dimension corresponds to intraday hours, while the spatial dimension corresponds to the bidding zones. The figure is constructed from hourly residuals of the NSTM–LASSO-3 final model (with post-processing error correcion) across all bidding zones and hours of the day, resulting in a $(24K \times 24K)$ correlation matrix. It is organized into hour blocks ($24\times24$), where each block represents zone-by-zone correlations across the $K$ bidding zones.
    }
\label{fig:hourzone-residuals_train}
\end{figure}

\clearpage
\vspace{-5mm} 
\nocite{*} 
\bibliographystyle{unsrtnat1}

\bibliography{bibliography} 

\end{document}